\input harvmac
\input amssym.def
\input amssym.tex

\def\^{{\wedge}}
\def\*{{\star}}
\def\bar{\overline}
\def\rk{\hbox{rank }}

\def\BC{{\bf C}}
\def\BP{{\bf P}}
\def\BR{{\bf R}}
\def\BZ{{\bf Z}}
\def\CD{{\cal D}}

\def\CI{{\cal I}}
\def\CL{{\cal L}}
\def\CM{{\cal M}}
\def\CN{{\cal N}}
\def\CO{{\cal O}}
\def\CV{{\cal V}}

\noblackbox

\def\urlfont{\hyphenpenalty=10000 \hyphenchar\tentt='057 \tt}

\newbox\tmpbox\setbox\tmpbox\hbox{\abstractfont PUPT-2081}
\Title{\vbox{\baselineskip12pt\hbox{\hss hep-th/0304115}
\hbox{PUPT-2081}}}
{\vbox{
\centerline{Residues and World-Sheet Instantons}}}
\smallskip
\centerline{Chris Beasley}
\smallskip
\centerline{\it{Joseph Henry Laboratories, Princeton University}}
\centerline{\it{Princeton, New Jersey 08544}}
\medskip
\centerline{and}
\medskip
\centerline{Edward Witten}
\smallskip
\centerline{\it{School of Natural Sciences, Institute for Advanced Studies}}
\centerline{\it{Princeton, New Jersey 08540}}
\bigskip\bigskip

We reconsider the question of which Calabi-Yau compactifications of
the heterotic string are stable under world-sheet instanton
corrections to the effective space-time superpotential.  For instance,
compactifications described by $(0,2)$ linear sigma models are
believed to be stable, suggesting a remarkable cancellation
among the instanton effects in these theories.  Here, we show that this
cancellation follows directly from a residue theorem, whose proof
relies only upon the right-moving world-sheet supersymmetries and
suitable compactness properties of the $(0,2)$ linear sigma model.
Our residue theorem also extends to a new class of ``half-linear''
sigma models.  Using these half-linear models, we show that heterotic
compactifications on the quintic hypersurface in $\BC \BP^4$ for which the
gauge bundle pulls back from a bundle on $\BC \BP^4$ are stable.  Finally,
we apply similar ideas to compute the superpotential contributions from
families of membrane instantons in M-theory compactifications on manifolds
of $G_2$ holonomy.

\Date{April 2003}

\lref\agntI{I. Antoniadis, E. Gava, K.~S. Narain, and T.~R. Taylor,
``Topological Amplitudes in String Theory,'' Nucl. Phys. {\bf B413}
(1994) 162--184, {\urlfont hep-th/9307158}.}

\lref\agntII{I. Antoniadis, E. Gava, K.~S. Narain, and T.~R. Taylor,
``Topological Amplitudes in Heterotic Superstring Theory,''Nucl. Phys.
{\bf B476} (1996) 133--174, {\urlfont hep-th/9604077}.}

\lref\basu{A. Basu and S. Sethi, ``World-sheet Stability of $(0,2)$
Linear Sigma Models,'' {\urlfont hep-th/0303066}.}

\lref\bbs{K. Becker, M. Becker, and A. Strominger, ``Fivebranes,
Membranes, and Non-Perturbative String Theory,'' Nucl. Phys. {\bf
B456} (1995) 130--152, {\urlfont hep-th/9507158}.}

\lref\bott{R. Bott, ``A Residue Formula for Holomorphic
Vector-Fields,'' J. Diff. Geom. {\bf 1} (1967) 311--330.}

\lref\bottandtu{R. Bott and L.~W. Tu, {\it Differential Forms in
Algebraic Topology}, Springer, New York, 1982.}

\lref\bsv{M. Bershadsky, V. Sadov, and C. Vafa, ``D-Branes and
Topological Field Theories,'' Nucl Phys. {\bf B463} (1996) 420--434,
{\urlfont hep-th/9511222}.}

\lref\carliebI{J.~B. Carrell and D.~I. Lieberman, ``Holomorphic Vector
Fields and Kaehler Manifolds,'' Invent. Math. {\bf 21} (1973) 303--309.}

\lref\carliebII{J.~B. Carrell and D.~I. Lieberman, ``Vector Fields and
Chern Numbers,'' Math. Ann. {\bf 225} (1977) 263--273.}

\lref\cmr{S. Cordes, G. Moore, and S. Ramgoolam, ``Lectures on 2D
Yang-Mills Theory, Equivariant Cohomology, and Topological Field
Theories,'' Nucl. Phys. Proc. Suppl. {\bf 41} (1995) 184--244,
{\urlfont hep-th/9411210}.}

\lref\distler{J. Distler, ``Resurrecting $(2,0)$ Compactifications,''
Phys. Lett. {\bf B188} (1987) 431--436.}

\lref\distnotes{J. Distler, ``Notes on $(0,2)$ Superconformal Field
Theories,'' {\urlfont hep-th/9502012}.}

\lref\distgreene{J. Distler and B. Greene, ``Aspects of $(2,0)$ String
Compactifications, Nucl. Phys. {\bf B304} (1988) 1--62.}

\lref\donald{S.~K. Donaldson and R.~P. Thomas, ``Gauge Theory in
Higher Dimensions,'' in {\it The Geometric Universe: Science,
Geometry, and the Work of Roger Penrose}, Ed. S.~A. Huggett et. al.,
Oxford University Press, Oxford, 1998.}

\lref\eguchi{T. Eguchi and S.~K. Yang, ``N=2 Superconformal Models as
Topological Field Theories,'' Mod. Phys. Lett. {\bf A5} (1990)
1693--1701.}

\lref\guk{S. Gukov, ``Solitons, Superpotentials, and Calibrations,''
Nucl. Phys. {\bf B574} (2000) 169--188, {\urlfont hep-th/9911011}.}

\lref\hl{R. Harvey and H.~B. Lawson, ``Calibrated Geometries,'' Acta
Math. {\bf 148} (1982) 47--157.}

\lref\hm{J.~A. Harvey and G. Moore, ``Superpotentials and
Membrane Instantons,'' {\urlfont hep-th/9907026}.}

\lref\instantons{M. Dine, N. Seiberg, X.~G. Wen, and E. Witten,
``Nonperturbative Effects on the String Worldsheet, I, II,''
Nucl. Phys. {\bf B278} (1986) 769--789, Nucl. Phys. {\bf B289} (1987)
319--363.}

\lref\gh{P. Griffiths and J. Harris, {\it Principles of Algebraic
Geometry}, John Wiley and Sons, Inc., New York, 1978.}

\lref\joyce{D.~D. Joyce, {\it Compact Manifolds with Special
Holonomy}, Oxford Univ. Press, Oxford, 2000.}

\lref\lawson{H.~B. Lawson, Jr., {\it Minimal Varieties in Real and
Complex Geometry}, S\'eminaire de Math\'ematiques Sup\'erieures
Universit\'e de Montr\'eal, 1974.}

\lref\leung{N.~C. Leung, ``Topological Quantum Field Theory for
Calabi-Yau Threefolds and $G_2$ Manifolds,'' {\urlfont math.DG/0208124}.}

\lref\liu{K. Liu, ``Holomorphic Equivariant Cohomology,''
Math. Ann. {\bf 303} (1995) 125--148.}

\lref\losev{A. Losev, N. Nekrasov, and S. Shatashvili, ``The Freckled
Instantons,'' {\urlfont hep-th/9908204}.}

\lref\morples{D.~R. Morrison and M.~R. Plesser, ``Summing the
Instantons: Quantum Cohomology and Mirror Symmetry in Toric
Varieties,'' Nucl. Phys. {\bf B440} (1995) 279--354,
{\urlfont hep-th/9412236}.}

\lref\mirror{E. Witten, ``Mirror Manifolds and Topological Field
Theory,'' in {\it Essays on Mirror Manifolds}, ed. S.-T. Yau, International
Press (1992) 120--160, {\urlfont hep-th/9112056}.}

\lref\silver{E. Silverstein and E. Witten, ``Criteria For Conformal
Invariance of $(0,2)$ Models,'' Nucl. Phys. {\bf B444} (1995) 161--190,
 {\urlfont hep-th/9503212}.}

\lref\tsikh{A.~K. Tsikh, {\it Multidimensional Residues and Their
Applications}, Trans. E.~J.~F. Primrose, American Mathematical
Society, Providence, R.~I., 1992.}

\lref\hettypei{J. Polchinski and E. Witten, ``Evidence for Heterotic-Type
I String Duality,'' Nucl. Phys. {\bf B460} 525--540, {\urlfont
hep-th/9510169}.}

\lref\pennI{E.~I. Buchbinder, R. Donagi, and B.~A. Ovrut,
``Superpotentials for Vector Bundle Moduli,'' {\urlfont
hep-th/0205190}.}

\lref\pennII{E.~I. Buchbinder, R. Donagi, and B.~A. Ovrut, ``Vector
Bundle Moduli Superpotentials in Heterotic Superstrings and
M-Theory,'' JHEP {\bf 0207} (2002) 066, {\urlfont hep-th/0206203}.}

\lref\branes{E. Witten, ``Branes and the Dynamics of QCD,'' Nucl. Phys.
{\bf B507} (1997) 658--690, {\urlfont hep-th/9706109}.}

\lref\phases{E. Witten, ``Phases of N=2 Theories in Two-Dimensions,''
Nucl. Phys. {\bf B403} (1993) 159--222, {\urlfont hep-th/9301042}.}

\lref\dinst{E. Witten, ``World-Sheet Corrections via D-Instantons,''
JHEP {\bf 0002} (2000) 030, {\urlfont hep-th/9907041}.}

\lref\simons{J. Simons, ``Minimal Varieties of Riemannian Manifolds,''
Ann. of Math. {\bf 88} (1968) 62--105.}

\lref\topfield{E. Witten, ``Topological Quantum Field Theory,''
Commun. Math. Phys. {\bf 117} (1988) 353--386.}

\lref\topsig{E. Witten, ``Topological Sigma Models,''
Commun. Math. Phys. {\bf 118} (1988) 411--449.}

\lref\sumor{E. Witten, ``Supersymmetry and Morse Theory,'' J. Differential 
Geometry {\bf 17} (1982) 661--692.}

\newsec{Introduction}

String theory backgrounds which preserve only $\CN=1$ supersymmetry in
four dimensions are of great interest both from a theoretical and a
phenomenological perspective.  A textbook way to obtain such
a background is to compactify either the $E_8 \! \times \! E_8$ or
$Spin(32)/\BZ_2$ heterotic string on a Calabi-Yau threefold $X$ with
a stable, holomorphic gauge bundle $E$.  One might suppose that these
compactifications, which admit a completely perturbative string
description, would be a natural starting point from which to study
the moduli space of $\CN=1$ backgrounds of string theory.

However, in fact we know very little about which pairs $(X,E)$ give
rise to consistent heterotic backgrounds, even in string perturbation
theory.  The issue, of course, is that models described by generic $X$
and $E$, even though they may satisfy the classical equations of
motion to all orders in $\alpha'$, are destabilized
non-perturbatively by world-sheet instantons \instantons.  These
instantons, arising from world-sheets which wrap rational
(i.e. holomorphic, genus zero) curves in $X$, can each contribute to a
background superpotential $W$ which lifts the K\"ahler moduli of $X$
and generates a cosmological constant.  So one might think that the
only stable $\CN=1$ heterotic compactifications would arise from very
special choices of $X$ and $E$ --- for instance corresponding to
world-sheet theories with $(2,2)$ supersymmetry or the $(0,2)$ models
studied by Distler and Greene \refs{\distler,\distgreene} --- for which
{\it each} world-sheet instanton simply cannot contribute to $W$.

In this light, the result of \silver\ that there are no non-perturbative
contributions to $W$ that destabilize compactifications described by
$(0,2)$ linear sigma models \refs{\phases, \distnotes} is somewhat
surprising.  This result does not rely upon any consideration of
world-sheet instantons and instead follows from simple facts about
the linear sigma model.  One simply observes that $W$ must always be a
holomorphic section of a complex line-bundle of strictly negative
curvature over the moduli space of the low-energy effective theory,
which is naturally a compact K\"ahler manifold in the case of a linear
sigma model.  The compactness of the moduli space implies that
$W$ must have a pole somewhere on the moduli space or else vanish
identically.  However, the linear sigma model, being a two-dimensional,
super-renormalizable gauge theory, can only become singular when the
target space becomes non-compact, as some bosonic field develops a
dangerous, unsuppressed zero-mode.  In computing the linear sigma
model correlators which describe the couplings of gauge-singlet fields
in the effective theory and so probe for a background $W$, one finds
that, after suitably twisting the model, no boson has a dangerous
zero-mode.  So $W$ has no poles on the moduli space and thus vanishes.

Now, Calabi-Yau compactifications which are described by $(0,2)$
linear sigma models are certainly not generic --- but nor are they so
special that each world-sheet instanton simply does not contribute to
$W$.  So from the world-sheet perspective, the stability of $(0,2)$
linear sigma models implies in these compactifications a remarkable
cancellation among the contributions to $W$ from world-sheet
instantons wrapped on rational curves in each homology class of $X$.

For instance, the analysis of \silver\ was applied in most detail to
the simple case that $X$ is a quintic hypersurface in $\BC \BP^4$
and $E$ is a deformation of the holomorphic tangent bundle $TX$,
corresponding to a deformation off the locus of $(2,2)$ supersymmetric
world-sheet theories.  In this case, the linear sigma model result
implies that, contrary to one's naive expectation, the world-sheet
instanton contributions to $W$ from the $2 875$ lines on the generic
quintic sum to zero.

Our main goal in this paper is to understand, from the world-sheet
perspective, the source of this remarkable cancellation among
instantons.  In the process, we will introduce a new $(0,2)$ ``half-linear''
sigma model and show that heterotic compactifications
described by these models form another class of stable $\CN=1$ string
backgrounds.  For instance, using the half-linear model we show that
heterotic compactifications on the quintic hypersurface in $\BC \BP^4$ for
which the gauge bundle pulls back from a bundle on $\BC \BP^4$ are stable.

More generally, just as for the linear models, the half-linear
models can be used to describe compactification on any Calabi-Yau
threefold $X$ which is a complete-intersection in a compact toric
variety $Y$.  However, in the half-linear models the bundle $E$ on
$X$ is now any stable, holomorphic bundle which pulls back from a
bundle on $Y$.  In particular, $E$ need not be a ``monad'' bundle
on $X$, the sort\foot{Technically, a monad bundle is one which
admits a description as the cohomology of a complex $A \rightarrow
B \rightarrow C$ of three bundles $A$, $B$, and $C$ on $X$.} most
naturally described in the linear sigma model.  Conversely,
however, there are also monad bundles on $X$ (including obvious
ones such as its tangent bundle) that do not pull back (at least
in any obvious way) from a holomorphic bundle on $Y$.  So we will
also develop a version of the vanishing argument adapted to linear
models and monad bundles on $X$.

\subsec{A Brief Sketch of the Idea}

Our essential idea can be motivated by considering the actual form
of the instanton contributions to $W$ in the simple case that the
string world-sheet wraps once about an isolated rational curve $C$ embedded
in $X$.  Actually, the most direct and elegant way \refs{\bbs, \dinst}
in this case to derive the instanton contribution to $W$ is to evaluate the
partition function of the worldvolume theory on a single $D1$-brane
wrapped on $C$ in the Type I theory, which is the dual description
\hettypei\ of a world-sheet instanton in the $Spin(32)/\BZ_2$
heterotic theory\foot{As explained in \dinst, the derivation of $W$
from the Type I theory most directly applies to the $Spin(32)/\BZ_2$
heterotic theory, but holomorphy and gauge-invariance allow us to
interpret the answer for the $E_8 \! \times \! E_8$ heterotic theory
as well.}.  Holomorphy allows us to evaluate this partition function
at one-loop, so the instanton contribution to $W$ from $C$ is just
\eqn\wcinst{
W(C) = \exp{\left( -{{A(C)} \over {2 \pi \alpha'}} + i \int_C B
\right)} \, {{{\hbox{Pfaff}'(\CD_F)}} \over {\sqrt{\det'(\CD_B)}}}
\,.}

Here the exponential factor in $W(C)$ represents the classical action
of the $D1$-brane.  We have written this action in heterotic units,
so that $A(C)$ is the area of $C$ in the heterotic string metric on $X$,
$\alpha'$ is the heterotic string tension, and $B$ is the heterotic
$B$-field.

The other factor in $W(C)$ arises from the one-loop
integral over the fluctuations of the bosons and fermions living on
the worldvolume of the $D1$-brane.  $\CD_B$ and $\CD_F$ are thus the
respective kinetic operators of the worldvolume bosons and fermions,
and the ``prime'' in $\det'(\CD_B)$ and $\hbox{Pfaff}'(\CD_F)$
indicates that these expressions are to be evaluated only after
omitting the zero-modes associated to the bulk symmetries which are
broken by the $D1$-brane.  Four bosonic zero-modes associated to the
broken translational symmetries in $\BR^4$ and two right-moving
fermionic zero-modes associated to the broken supersymmetries arise in
this fashion.

The complex structure moduli of $X$ and $E$ are described by chiral
superfields in the low-energy, effective $\CN=1$ theory, and $W(C)$
must depend holomorphically on these fields.  Unfortunately, our
simple expression \wcinst\ for $W(C)$ is not manifestly holomorphic.
To get a manifestly holomorphic expression for $W(C)$, we must use
the fact that the two supersymmetries left unbroken by the $D1$-brane
imply a cancellation between the contributions of the right-moving
fermionic modes to $\hbox{Pfaff}'(\CD_F)$ and the contributions of the
right-moving bosonic modes to $\det'(\CD_B)$.

To make this cancellation explicit, we write $W(C)$ solely
in terms of the left-moving bosonic and fermionic modes.  By convention,
the kinetic operator of a left-moving fermion on $C$ will be a $\overline
\partial$ operator, while the kinetic operator for a right-moving
fermion will be a $\partial$ operator.  Thus, since the left-moving
worldvolume fermions transform as sections of the left-moving
spin bundle\foot{Here $\CO(n)$ is the usual notation for
the complex line-bundle of degree $n$ on projective space.  In
particular, $\CO$ is the trivial complex bundle of rank one.}
$S_- = \CO(-1)$ on $C$ tensored with the gauge bundle $E$ as
restricted to $C$, their contribution to $\hbox{Pfaff}'(\CD_F)$ is
just the Pfaffian of the $\overline \partial$ operator coupled to
$E \otimes \CO(-1) \equiv E(-1)$, which we denote
$\overline \partial_{E(-1)}$.

Similarly, in the formula \wcinst\ for $W(C)$, we have written the boson
kinetic operator $\CD_B$ as a real operator acting on the eight real
bosons representing the normal directions to $C$ in $\BR^4 \times X$.
Since $C$, $X$, and $\BR^4$ all have complex structures, we can
equally well group the eight real bosons into four complex bosons
taking values in the complex normal bundle $N$ to $C$ in $\BR^4
\times X$.  When $C$ is isolated in $X$, $N$ is isomorphic to $\CO
\oplus \CO \oplus \CO(-1) \oplus \CO(-1)$, the first two summands
representing the normal directions in $\BR^4$ and the last two summands
representing the normal directions in $X$.  Thus, the contribution of
the non-zero left-moving bosonic modes to $\det'(\CD_B)$ just
arises from the $\overline \partial$ operator on $C$ coupled
to the normal bundle $N$.

So, cancelling out the right-moving modes from $W(C)$ in \wcinst, we
have
\eqn\holwcinst{
W(C) = \exp{\left( -{{A(C)} \over {2 \pi \alpha'}} + i \int_C
B \right)} \, {{\hbox{Pfaff}\left(\overline \partial_{E(-1)}(C)\right)} \over
{\left(\det'{\overline \partial_\CO}\right)^2 \left(\det{\overline
\partial_{\CO(-1)}(C)}\right)^2}} \,.}
This expression for $W(C)$ is now manifestly holomorphic.  Specifying
a $\overline \partial$ operator on either $X$ or $E$ is equivalent to
specifying its complex structure, so the operators
$\overline \partial_{E(-1)}(C)$ and $\overline \partial_{\CO(-1)}(C)$
themselves depend holomorphically on the complex structure moduli of
$X$ and $E$.  We have also emphasized in \holwcinst\ that the way in
which the complex structure moduli of $X$ and $E$ appear in these
$\overline \partial$ operators itself depends upon which curve $C$ in
$X$ that the instanton wraps.  In fact, at least when $X$ is
elliptically fibered, one can derive very explicit expressions in
given examples for the dependence of $W(C)$ on the moduli of $E$ and
$X$ \refs{\pennI, \pennII}, although we will not be needing such
detailed expressions here.

Thus, in the case that $X$ is a generic quintic in $\BC \BP^4$ and $E$
is a deformation of $TX$, the vanishing of $W$ implies as a corollary
that, summing $W(C)$ over the $2875$ lines on $X$,
\eqn\corollary{
\sum_{i=1}^{2875} \; {{\exp{\Big( i \int_{C_i} \! B \Big)} \;
\hbox{Pfaff}\left(\overline \partial_{E(-1)}(C_i)\right)} \over
{\left(\det{\overline \partial_{\CO(-1)}(C_i)}\right)^2}} = 0\,.}
In this expression, we have dropped from $W(C)$ an overall factor of
$\exp{\Big({{-A(C)} \over {2 \pi \alpha'}}\Big)}$, which is constant for
curves on $X$ of given degree, and a factor of $(\det'{\overline
\partial_\CO})^{-2}$, which is simply constant.

One is very much also tempted to drop from \corollary\ the factor
of $\exp{(i \int_{C_i} \! B)}$, which is at least ``morally'' constant
on curves of given degree.  However, as reviewed in \dinst, because
only the product of $\exp{(i \int_{C_i} \! B)}$ and the fermion
Pfaffian is even well-defined, we must technically include in
\corollary\ this factor involving $B$ so that the full expression
makes sense.  Nevertheless, our interest in \corollary\ resides in the
holomorphic dependence of this formula on the complex structure moduli
of $X$ and $E$, and we will not dwell here on the subtleties of the
heterotic $B$-field.

At first sight, the formula \corollary\ might seem like an exotic
mathematical prediction derived only indirectly from the underlying
$(0,2)$ linear sigma model.  But in fact, this sort of formula has a
clear precedent from algebraic geometry, in the form of a residue
theorem.

To derive the simplest example of such a residue theorem, suppose that
$\omega$ is a meromorphic one-form on $\BC \BP^1$ with simple poles
at points $P_i$, $i=1,\ldots,N$.  Letting $z$ be a holomorphic coordinate
on $\BC \BP^1$, we can express $\omega$ as
\eqn\form{
\omega = {{g(z) dz} \over {f(z)}}\,,}
where $f$ and $g$ are polynomials in $z$, $f$ having non-degenerate
zeroes at the points $P_1, \ldots, P_N$.  Without loss, we assume
that $\omega$ does not have a pole at $z = \infty$, so that the degrees of
$f$ and $g$ satisfy
\eqn\degs{ \deg{g} \le \deg{f} - 2\,.}

As usual, we then define the residue of $\omega$ at each point $P_i$, denoted
$\hbox{Res}_{P_i}(\omega)$, by integrating $\omega$ over a small
contour $\gamma_i$ about the point $P_i$,
\eqn\onedres{
\hbox{Res}_{P_i}(\omega) = {1 \over {2 \pi i}} \, \oint_{\gamma_i}
\omega \,= {{g(P_i)} \over {\partial f / \partial z (P_i)}} \,.}
We now obtain a residue theorem simply by considering the sum of contours
\eqn\bigg{ \Gamma = \gamma_1 + \cdots + \gamma_N\,.}
Since $\Gamma$ is contractible, we have
\eqn\resthm{
0 = {1 \over {2 \pi i}} \, \oint_\Gamma \omega = \sum_{i=1}^N
\hbox{Res}_{P_i}(\omega) = \sum_{i=1}^N {{g(P_i)} \over {\partial f /
\partial z (P_i)}} \,.}
So the residue theorem simply states that the sum of the residues of
$\omega$ is zero.

Comparing \corollary\ to \resthm, we can already see a vague
similarity between these two formulae, with the Pfaffian in
\corollary\ being a holomorphic function analogous to $g$ in \resthm,
and the bosonic determinant in \corollary\ being analogous to
$\partial f / \partial z$ in \resthm.  Our main goal in this paper is
to make the correspondence between these formulae precise, showing
directly that the instanton contributions to $W$ vanish in
suitable models due to an infinite-dimensional generalization
of the simple one-dimensional residue theorem above.

\subsec{The Plan of the Paper}

Our plan for the paper is as follows.  In Section 2, we start by
generalizing the one-dimensional residue theorem to finitely many
dimensions.  Although standard mathematical approaches exist for
studying multi-dimensional residues, as for instance in \refs{\gh, \tsikh},
we will take a more physical approach by studying a certain
supersymmetric, finite-dimensional integral.  This integral is a
natural abstraction of the path-integral over the right-moving
world-sheet fields on the heterotic string, and from it we easily
prove a very general, multi-dimensional residue theorem.

At the end of Section 2, we also describe precisely how the partition
function of the worldvolume theory on a supersymmetric
$D1$-brane can be interpreted as a residue.  Unfortunately, although
the $D1$-brane formalism provides a very elegant description of the
superpotential contribution from any single instanton, the world-sheet
description of the heterotic string turns out to be better for proving
vanishing results for the sums of these instanton contributions.

So in Section 3, we apply our analysis from Section 2 to the
heterotic world-sheet theory itself.  This analysis most directly
generalizes to the ``half-linear'' class of heterotic
compactifications, for which $X$ is a complete-intersection in a
compact toric variety $Y$ and the gauge bundle $E$ on $X$ pulls
back from a bundle on $Y$.  For these compactifications, the
vanishing of the instanton contributions to $W$ follows from
essentially the same argument as we use in Section 2 to deduce the
multi-dimensional residue theorems.  We also show how this
argument can be applied to the $(0,2)$ linear sigma models to
prove directly formulae such as \corollary.

Very recently, Basu and Sethi \basu\ have also given another
argument for the stability of $(0,2)$ linear sigma models.  Their
argument focuses on showing the absence of corrections to the
world-sheet superpotential.

Finally, in Section 4 we consider the $\CN=1$ compactification of
M-theory on a manifold $X$ of $G_2$ holonomy.  Using ideas very similar to
those in Sections 2 and 3, we extend the results of \hm\ by computing
the superpotential contribution from membranes which wrap a continuous
family of supersymmetric three-cycles in $X$.

\newsec{Residues and Supersymmetry}

Rather than trying to generalize immediately from the
one-dimensional residue theorem to an infinite-dimensional residue
theorem which is applicable to the heterotic string, we will warm
up with the simpler generalization to residue theorems in only a
finite number of dimensions.  Our strategy is to consider a
finite-dimensional, supersymmetric integral on an arbitrary
compact, complex manifold $M$. The finite-dimensional residue
theorem then follows from the supersymmetry, which allows us to
localize the integral to a sum of terms generalizing the
one-dimensional residues, and from the compactness of $M$, which
leads to the vanishing of the integral and hence the sum.  After
we obtain this result, we will indicate some easy generalizations
of it which also have relevance to the heterotic models we
introduce in Section 3.  Finally, we describe precisely how the
partition function of the worldvolume theory on supersymmetric
$D1$-brane can be interpreted as a residue.

Standard mathematical approaches to multi-dimensional residues and
residue theorems can be found in \gh\ and \tsikh.  Mathematical
discussions somewhat more related to our approach via supersymmetry
are given in \carliebI, \carliebII, and \liu.

\subsec{A Finite-Dimensional Integral}

We now introduce the finite-dimensional, supersymmetric integral that
is central to our study of residues and which serves as a model for
the path-integral over the world-sheet fields of the heterotic
string.  Since the supersymmetry in our integral is essential, we will
begin by specifying how it acts on the variables of integration.

As mentioned above, we perform the integral over a compact, complex
manifold $M$, having (complex) dimension $n$.  So the bosonic
variables of integration will be the local holomorphic and
anti-holomorphic coordinates $z^i$ and $z^{\bar i} \equiv
\overline{z^i}$ on $M$.

We also introduce a set of anti-commuting, fermionic coordinates
$\theta^{\bar i}$ and $\chi^\alpha$.  Here the fermions $\theta^{\bar
i}$ transform as coordinates on the anti-holomorphic tangent bundle
$\overline{TM}$, and the fermions $\chi^\alpha$ transform as
coordinates on a holomorphic vector bundle $V$, of rank $r$, over
$M$.  The bundle $V$ is completely arbitrary and should be
considered, like $M$, as part of the defining data for our integral.

Besides specifying $V$ itself, we must now also choose a global
holomorphic section $s$ of $V$.  We need this section $s$ simply to
define an interesting supersymmetry transformation for the fermions
$\chi^\alpha$, since none of the other variables of integration
have anything to do with $V$.  So under the supersymmetry $\delta$,
the bosonic and fermionic variables transform as
\eqn\littleQ{\matrix{
&\eqalign{
&\delta z^i = 0\,,\cr
&\delta \chi^\alpha = s^\alpha \,,\cr}
&\eqalign{
&\delta z^{\bar i} = \theta^{\bar i} \,,\cr
&\delta \theta^{\bar i} = 0 \,.\cr}\cr}}
Note that since $s$ is holomorphic, $\delta^2 = 0$, the most important
property of $\delta$.

The supersymmetric integral which we consider takes the general form
\eqn\bigz{
Z = \int_M g \, d \mu \, \exp{(-t \, S)} \,,}
where $t$ is a positive real parameter representing the
``coupling constant'' for $Z$, $S$ is a finite-dimensional ``action''
which we will soon present, and
\eqn\meas{ g \, d \mu \equiv g(z) \, d^n z \, d^n \overline z \,
d^n \theta \, d^r \chi }
is the measure.  Locally, $g$ is a function which represents the
particular choice of measure for $Z$, and to ensure that the measure
respects the supersymmetry, $g$ must be holomorphic.

The fact that we have to worry about the measure for $Z$ may
seem slightly odd, since in many supersymmetric integrals, one can
make a canonical choice of measure (up to normalization).  The
point is that, under any change of variables, the resulting Jacobians
for the bosonic variables are cancelled by the fermionic Jacobians for
their superpartners.

In the case of $d \mu$ above, such a cancellation occurs
between the anti-holomorphic bosons $z^{\bar i}$ and their
superpartners $\theta^{\bar i}$.  So the factor $d^n \overline z \,
d^n \theta$ appearing in $d \mu$ indeed represents a canonical choice
of measure for these variables.

On the other hand, the bosonic variables $z^i$ and the fermionic
variables $\chi^\alpha$ are unrelated by supersymmetry, which
means that we really must choose the factor $g(z) \, d^n z \, d^r
\chi$ appearing in \meas.  Globally, $g$ is not a function but
transforms as a holomorphic section of the line-bundle $\Omega^n_M
\otimes \wedge^r V$ on $M$, where $\Omega^n_M$ denotes as usual
the canonical bundle of holomorphic $n$-forms on $M$ and $\wedge^r
V$ is the top exterior power of $V$.  Since we generally have no
preferred choice of such a section, we must interpret our choice
of $g$ as another part of the input data needed to specify $Z$.

We must, of course, also specify the action $S$ for the integrand of $Z$.
We first choose a positive-definite, hermitian metric $h_{\bar \alpha
\alpha}$ on $V$.  Then we consider a $\delta$-trivial action,
\eqn\sone{ S = \delta \left( h_{\bar \alpha \alpha} \,
s^{\bar \alpha} \chi^\alpha \right) \,,}
or expanding,
\eqn\stwo{ S = h_{\bar \alpha \alpha} s^{\bar \alpha} s^\alpha
+ h_{\bar \alpha \alpha} D_{\bar j} s^{\bar \alpha} \,
\theta^{\bar j} \chi^\alpha \,.}
Here $D_{\bar j}$ is the covariant derivative associated to the canonical
connection arising from the metric $h_{\bar \alpha \alpha}$ on $V$.
Recall that the canonical connection \gh\ is the unique connection on $V$
for which $h_{\bar \alpha \alpha}$ is covariantly constant and for
which $D_{\bar j} = \partial_{\bar j}$ when acting on a holomorphic
frame of $V$.

One easy consequence of the fact that $S$ is $\delta$-trivial is that
$S$ is obviously supersymmetric, $\delta S = 0$.  A deeper
consequence of the fact that $S$ is $\delta$-trivial is that the integral
$Z$ is formally independent of the real parameter $t$ and the metric
$h_{\bar \alpha \alpha}$ on $V$ which we introduced.  For instance,
the invariance of $Z$ under changes in $t$ is derived by first
observing that
\eqn\dt{{{d Z} \over {d t}} = - \int_M g \, d \mu \, S \exp{(-t \,
S)} = - \langle S \rangle \,.}
However, if $\CO$ is any function of the variables $z^i$, $z^{\bar
i}$, $\chi^\alpha$, and $\theta^{\bar i}$, then
\eqn\decouple{ \langle \delta \CO \rangle =
\int_M g \, d \mu \, \delta \CO \, \exp{(-t \, S)} = 0\,,}
which in the language of topological field theory is the decoupling of
BRST-trivial observables \refs{\topfield, \cmr}.  Since the action $S$
is of the form $\delta \CO$, we deduce immediately that $d Z / d t = 0$.  The
invariance of $Z$ under deformations of the metric $h_{\bar \alpha
\alpha}$ follows by the same argument.

Finally, we observe that $S$ is invariant under a ghost number
symmetry, under which the anti-commuting variables $\chi^\alpha$ and
$\theta^{\bar i}$ carry charges $-1$ and $+1$ respectively, and
$\delta$ itself carries charge $+1$.  Since the measure $d \mu$ thus
carries ghost number
\eqn\ghostnum{ \dim M - \rk V = n - r\,,}
$Z$ vanishes identically unless $n = r$.  So, if we wish to use $Z$ to
prove a residue theorem, we must assume that $\dim M = \rk V$.

\subsec{A Residue Theorem}

As is familiar from the study of other topological models, we can
prove an interesting theorem by using the fact that $Z$ is
independent of $t$ and then evaluating $Z$ for $t \rightarrow
\infty$ and $t = 0$.  Sometimes, a formal statement such as ``$Z$
is independent of $t$'' could fail to hold if the convergence of $Z$ were
sufficiently poor.  See \silver\ for a nice demonstration of such a
failure in the context of the linear sigma model.  However, because
here $Z$ is an integral over a compact manifold $M$, the convergence
of $Z$ is assured, even when $t=0$, and there are no difficulties with
the formal statements above.

Evaluating $Z$ when $t=0$ is easy.  Then
\eqn\tzero{ Z = \int_M g \, d \mu \; 1 \,= 0\,,}
since neither $\chi^\alpha$ nor $\theta^{\bar i}$ appear in the
integrand above.

Evaluating $Z$ for $t \rightarrow \infty$, we see from the action $S$
in \stwo\ that only points in a neighborhood of the vanishing locus $L$
of the section $s$ contribute to $Z$.  In general, $L$ will consist of
several disconnected components $C$, and $Z$ must have an expression
\eqn\tinf{ Z = \sum_{C \subset L} \, Z(C) \,,}
where $Z(C)$ denotes the local contribution to $Z$ from the component
$C$.  So \tzero\ and \tinf\ imply as a very general vanishing theorem
that
\eqn\vanishing{ \sum_{C \subset L} \, Z(C) = 0 \,.}

The power of this approach is that the vanishing theorem \vanishing\
does not rely on any specific behavior of the section $s$ of $V$.  In
the simplest case, $s$ has simple zeroes on a set of isolated points
of $M$.  But we can equally well consider the case that $s$ has
degenerate zeroes at some points, or even that $s$ vanishes over
some components of positive dimension.  In order to translate
\vanishing\ into a more explicit formula, along the lines of the
one-dimensional residue theorem \resthm, we must simply evaluate the
expression $Z(C)$ for each case.

\bigskip\noindent{\it Multi-dimensional residues}\smallskip\nobreak

To make contact with the one-dimensional residue theorem \resthm, we
will consider at first only the easiest case that $s$ vanishes in a
non-degenerate fasion on a set of isolated points $P$ of $M$.

Recall that the requirement that $s$ vanish non-degenerately at a
point $P$ is simply the condition that the Jacobian $\det{(ds)}$
be non-vanishing at $P$,
\eqn\jac{
\det{(ds)}(P) = \det{\Big( {{\partial(s^1,\cdots,s^n)} \over
{\partial(z^1,\cdots,z^n)}} \Big)}(P) \neq 0.}
In this case, the contribution $Z(P)$ from $P$ can be evaluated
exactly using the Gaussian approximation to $Z$ near this point,
and we easily see that\foot{In the following, we suppress
overall factors of $\pi$ that arise from the Gaussian integration.}
\eqn\zp{ Z(P) = {{g(P)} \over {\det{(ds)}(P)}} \,.}
Thus, the vanishing result \vanishing\ becomes
\eqn\resnthm{ \sum_{P \in L} \, {{g(P)} \over {\det{(ds)}(P)}} = 0\,.}
This expression represents a natural generalization of the
one-dimensional residue theorem \resthm.

To sharpen the correspondence between the formula \resnthm\ and a
multi-dimensional residue theorem, we consider the particular case
that the bundle $V$ is a direct sum of $n$ line bundles,
\eqn\V{ V = \CO(D_1) \oplus \cdots \oplus \CO(D_n) \,,}
which are associated to $n$ irreducible, effective divisors
$D_1,\ldots,D_n$ intersecting transversely at isolated points $P$ in
$M$.

To describe the appropriate section $s$ of $V$ for this case, we note
that each divisor $D_i$ is determined as the vanishing locus
of a holomorphic section $s_i$ of the line-bundle $\CO(D_i)$.  Then we
simply take $s$ to be the direct sum of the $s_i$, so that $s$ has
components
\eqn\s{ s = (s_1,\ldots,s_n) \,.}
We note that the section $s$ vanishes non-degenerately at each point
$P \in D_1 \cap \cdots \cap D_n$, so our simple expression for $Z(P)$
in \zp\ is valid.

In this case, we can now give a very nice geometric interpretation
of the local contribution $Z(P)$ from each point $P \in D_1 \cap
\cdots \cap D_n$.  Near $P$, we can trivialize all the line
bundles $\CO(D_i)$ as well as the canonical bundle of $M$.  Upon
doing so, we can regard $g$ as an ordinary holomorphic function
that is nonzero at $P$, and the $s_i$ as holomorphic functions
that vanish on $D_i$.  Now we can define a meromorphic $n$-form
$\omega$ that generalizes the one-dimensional expression \form,
\eqn\formn{ \omega = {{g \, dz^1 \^ \cdots \^ dz^n} \over {s_1
\cdots s_n}}\,.}

Given the meromorphic $n$-form $\omega$, and a real $n$-cycle
$\gamma$ that links in a suitable way the locus of its poles, we
can naturally define an $n$-dimensional residue
$\hbox{Res}_P(\omega)=(1/2\pi i)^n\int_\gamma\omega$ that will
generalize the usual one-dimensional residue. We let $\gamma$ be
the real $n$-cycle determined by \eqn\ncycle{ |s_i| =
\epsilon\,,\quad i=1,\ldots,n\,,} where $\epsilon$ is a small
parameter. Technically, we must also orient $\gamma$, which we do
by the condition ${d( \arg s_1) \^ \cdots \^ d(\arg s_n) \ge 0}$.

On $\gamma$, $\omega$ is holomorphic, so we can define \eqn\ndres{
\hbox{Res}_{P}(\omega) = \left({1 \over {2 \pi i}}\right)^n \,
\int_\gamma \omega \,.} Since $d \omega = 0$ on a neighborhood of
$\gamma$, this definition only depends on the homology class of
$\gamma$ and in particular does not depend the parameter
$\epsilon$ above.

The residue $\hbox{Res}_P(\omega)$ can be then be evaluated by a
change of variables and the iterative application of Cauchy's theorem.
We find
\eqn\ndresform{
\hbox{Res}_P(\omega) = {{g(P)} \over {\det{(ds)}(P)}} \,,}
generalizing the one-dimensional expression in \onedres.  Of course,
$\hbox{Res}_P(\omega)$ agrees precisely with $Z(P)$ for the special
choices of $V$ and $s$ above, so our main result \vanishing\ is properly
interpreted as a generalized, multi-dimensional residue theorem.

\bigskip\noindent{\it A quick example}\smallskip\nobreak

Before proceeding further, we will give a quick example of the residue
theorem.

For our example, we take $M = \BC \BP^2$ and $V = TM$, the
holomorphic tangent bundle.  If we let $[X_0:X_1:X_2]$ be homogeneous
coordinates on $M$, then any holomorphic section $s$ of
$V$ takes the form
\eqn\ptwosec{ s = a_0 \, X_0 \, {\partial \over {\partial X_0}} +
a_1 \, X_1 \, {\partial \over {\partial X_1}} + a_2 \, X_2 \,
{\partial \over {\partial X_2}} \,,}
where $(a_0,a_1,a_2)$ are complex coefficients parametrizing $s$.
Because $[X_0:X_1:X_2]$ are only homogeneous coordinates, the coefficients
$(a_0,a_1,a_2)$ are only defined up to the addition of a multiple of
$(1,1,1)$, which describes the zero section of $V$.  If
$(a_0,a_1,a_2)$ are generic coefficients, then $s$ vanishes
non-degenerately at the three points ${P_1 = [1:0:0]}$, ${P_2 = [0:1:0]}$,
and ${P_3 = [0:0:1]}$ of $M$.

Since $V = TM$, the measure $d \mu$ is a section of the trivial
bundle $\CO = \Omega^n_M \otimes \wedge^n T M$.  Consequently, in
this example we do have a canonical measure for $Z$ and $g$ is a
constant.

Now in the patch where $X_0 \neq 0$, with local coordinates
$(z^1,z^2)$, $s$ takes the form
\eqn\patchone{ s = (a_1 - a_0) \, z^1 {\partial \over {\partial z^1}}
+ (a_2 - a_0) \, z^2 {\partial \over {\partial z^2}} \,,}
and so the residual contribution from $P_1$ to $Z$ is
\eqn\zpone{ Z(P_1) = {1 \over {(a_1 - a_0) (a_2 - a_0)}}\,.}
Similar contributions from the points
$P_2$ and $P_3$ are
\eqn\zpetc{ Z(P_2) = {1 \over {(a_0 - a_1) (a_2 - a_1)}}\,,\quad
Z(P_3) = {1 \over {(a_0 - a_2) (a_1 - a_2)}}\,.}
The residue theorem then simply states that $Z(P_1) + Z(P_2) + Z(P_3) = 0$,
as one can verify directly.

\subsec{Generalizations}

The ghost number symmetry preserved by $S$ implies that $Z$ trivially
vanishes unless $\rk V = \dim M$.  So if we wish to study bundles $V$
such that $\rk V \neq \dim M$, we should consider not $Z$ itself
but expectation values $\langle \CO \rangle$,
\eqn\expect{
\langle \CO \rangle = \int_M  g \, d \mu \, \CO \, \exp{(-t S)}\,,}
where $\CO$ is any function of $z^i$, $z^{\bar i}$, $\chi^\alpha$, and
$\theta^{\bar i}$ which satisfies $\delta \CO = 0$.  Of course, $\CO$
must also have ghost number $n - r$ if $\langle \CO \rangle$ is to be
any more interesting that $Z$ itself.

Globally, functions $\CO$ of $z^i$, $z^{\bar i}$, $\chi^\alpha$,
and $\theta^{\bar i}$ are elements of the complex \eqn\cplx{
\bigoplus_{(p,q)} \, A^{(0,q)}(M) \otimes \wedge^p V^*\,.} Here
$A^{(0,q)}(M)$ is the bundle of smooth $(0,q)$ forms on $M$, and
$V^*$ is the holomorphic bundle dual to $V$.  A function
homogeneous and $q^{th}$ order in $\theta^{\bar i}$ is a
$(0,q)$-form on $M$, while a function homogeneous and $p^{th}$
order in $\chi^\alpha$ is a section of $\wedge^pV^*$.  We will
often refer to an element of $A^{(0,q)}(M) \otimes \wedge^p V^*$
for fixed $(p,q)$ as having ``type'' $(p,q)$.

The supersymmetry transformation $\delta$ acts on elements of this
complex as \eqn\bigD{ D = \theta^{\bar i} {\partial \over
{\partial z^{\bar i}}} + s^\alpha {\partial \over {\partial
\chi^\alpha}}\,.} More intrinsically, we can identify $D$ with the
operator \eqn\bigDI{ D = \overline \partial + \iota_s\,,} where
$\overline \partial$ is the usual Dolbeault operator on $M$ and
$\iota_s$ acts on sections of $\wedge^p V^*$ by the interior
product with $s$.  The action of $D$ on this complex has certainly
been considered before in the mathematical literature, for
instance in \refs{\carliebI, \carliebII, \liu}, though mostly for
the case $V = TM$.

Since $\langle \delta \CO \rangle = 0$ for any $\CO$, the
interesting observables $\CO$ correspond to nontrivial elements of
the cohomology of $D$.  In general, what can we say about this
cohomology?

Without placing additional conditions on $M$, $V$, and $s$, in fact we
cannot say much\foot{However, see \liu\ for a nice discussion of
the easiest case that $V = TM$ and $s$ has zeroes at isolated points.
In this case, the cohomology of $D$ is isomorphic to $H^0(M, \CO /
\CI)$, where $\CI$ is the ideal sheaf associated to $s$.}.  Nonetheless,
we do have a systematic procedure to compute the $D$-cohomology, using a
spectral sequence (see \bottandtu\ for a clear introduction to spectral
sequences).

In physical terms, we want to solve the equation $\delta \CO = 0$, and
the spectral sequence is essentially a perturbative way to do this,
really by following one's nose.  So to construct an $\CO$ which
satisfies $\delta \CO = 0$, we start with an ``order-zero'' trial
solution $\CO^{(0)}$, of type $(p,q)$, which satisfies\foot{If we
wished, we could equally well start with $\CO^{(0)}$ satsifying
$\iota_s \CO^{(0)} = 0$ and reverse the roles of $\overline \partial$
and $\iota_s$ above.  We find it convenient to do this in Section 3.}
$\overline \partial \CO^{(0)} = 0$.  If $\CO^{(0)}$
also happens to satisfy $\iota_s \CO^{(0)} = 0$, then $\CO =
\CO^{(0)}$, but generally $\iota_s \CO^{(0)} \neq 0$.

To correct for this discrepancy, we then try to solve
\eqn\Oone{ \iota_s \CO^{(0)} + \overline \partial \CO^{(1)} = 0\,,}
to determine the ``first-order'' correction $\CO^{(1)}$.  We consider
$\CO^{(1)}$ as a correction to $\CO^{(0)}$ in a very definite sense,
since although $\CO^{(0)}$ is of type $(p,q)$, $\CO^{(1)}$ is of type
$(p-1,q-1)$.  Thus, if we continue to solve iteratively
\eqn\Op{ \iota_s \CO^{(n)} + \overline \partial \CO^{(n+1)} = 0\,,}
we will either find an obstruction, or the procedure will terminate
after a finite number of steps with $\CO = \CO^{(0)} + \CO^{(1)} +
\cdots$ satisfying $\delta \CO = 0$.  We will find this little
procedure useful when constructing heterotic models in Section 3.

What sort of results, analogous to the generalized residue theorem
\vanishing, do we then obtain by considering the expectations
$\langle \CO \rangle$ of nontrivial observables $\CO$?
Evaluating $\langle \CO \rangle$ at $t=0$ now yields
\eqn\Otzero{ \langle \CO \rangle = \int_M g \, d \mu \, \CO \,,}
which need not vanish if $\CO$ carries the proper ghost number.
Evaluating $\langle \CO \rangle$ in the limit $t \rightarrow \infty$,
we again see that $\langle \CO \rangle$ can be expressed as a sum of local
contributions from each of the components $C$ of the vanishing locus
$L$ of $s$,
\eqn\Otinfty{
\langle \CO \rangle = \sum_{C \subset L} \, \langle \CO \rangle(C) \,.}

So for instance, again in the case that $s$ vanishes non-degenerately
over isolated points $P$ of $M$ and $\CO$ has ghost number zero,
\eqn\Ozero{
\int_M g \, d \mu \, \CO = \sum_{P \in L} \, {{g(P) \, \CO(P)} \over
{\det{(ds)}(P)}} \,.}
In the above expression, we must interpret the integral over $M$ as
picking out the component of $\CO$ of type $(n,n)$ and the evaluation
at $P$ as picking out the component of $\CO$ of type $(0,0)$.

We can also consider the under-determined case, for which $\rk V <
\dim M$, as well as the over-determined case, for which $\rk V > \dim
M$.  In the under-determined case, the components $C$ of $L$ will
generically be complex submanifolds of dimension $n - r$ in $M$.  We
assume that $s$ vanishes in a non-degenerate fashion on each $C$,
which means that the Jacobian $\det{(ds|_N)}$ of $s$ with respect to the
normal directions to $C$ in $M$ is non-vanishing along $C$.  Then the
local contribution of $C$ to $\langle \CO \rangle$ is
\eqn\OC{ \langle \CO \rangle(C) = \int_C {{g \, d \mu \, \CO} \over
{\det{(ds|_N)}}} \,.}
In the above, $g \, d \mu \, / \det{(ds|_N)}$ determines an element
of $\Omega^{n-r}$ on $C$, and thus only the component of $\CO$ of
type $(0,n-r)$ now contributes to the integral over $C$.

As we shall see in Section 3, the case of direct relevance to the
heterotic string is actually the over-determined case, $\rk V >
\dim M$.  In this case, one might think that it is unnatural to
consider an $s$ that has zeroes.  However, to get a non-trivial
result in the over-determined case, we need a non-trivial $\CO$,
most simply an $\CO$ of degree $(r-n,0)$.  In any case, such an
$\CO$ will be present when we study the half-twisted heterotic
string.  ($r-n$ will be infinite, and $\CO$ will be the
exponential of a fermion bilinear.) When such an $\CO$ is present,
$s$ cannot be changed freely, since one must preserve the
condition $(\overline\partial+\iota_s)\CO=0$. In the presence of a
suitable $\CO$, it can be natural to have an $s$ that has zeroes
with non-trivial residues. For instance, if $s$ again vanishes
non-degenerately at an isolated point $P$ of $M$, now meaning that
the matrix $ds = \left({{\partial s^\alpha} / {\partial
z^i}}\right)$ has full rank at $P$, then the local contribution
from $P$ to $\langle \CO \rangle$ is \eqn\Oneg{ \langle \CO
\rangle(P) = \left( {{g \, d \mu \, \CO} \over {ds}} \right) (P)
\equiv \left( {{g \, \epsilon_{i_1 \cdots i_q} \epsilon^{\alpha_1
\cdots \alpha_p} \, \CO_{\alpha_{q+1} \cdots \alpha_p}} \over
{\partial_{i_1} s^{\alpha_1} \cdots
\partial_{i_q} s^{\alpha_q}}}\right) (P) \,.}
Evidently, in such an example with isolated zeroes of $s$, only the
component of type $(r-n,0)$ of $\CO$ contributes to
$\langle \CO \rangle(P)$.

\subsec{The $D1$-brane Partition Function as a Residue}

Our discussion of multi-dimensional residues now allows us to make
precise the manner in which the partition function of a supersymmetric
$D1$-brane can be interpreted as a residue.  We have already seen
in the Introduction a strong formal similarity between expressions
such as \corollary\ and \resthm\ which suggests this interpretation.
To check this idea, though, we must examine to what extent the
worldvolume theory on a supersymmetric $D1$-brane actually generalizes
our finite-dimensional model which produces the residues.

At first glance, one might be worried by the following fact.  If
we consider the bosonic action for a $D1$-brane which wraps an
arbitrary, not necessarily holomorphic, surface $\Sigma$ in the
Calabi-Yau threefold $X$, then this action is just the area
$A[\Sigma]$ of the surface.  So if the $D1$-brane action were
literally to be the obvious generalization of the action \stwo\ of
the finite-dimensional model, then $A[\Sigma]$ would have to admit
a representation as the norm-squared of a suitable holomorphic
section $s$ over the space $\CM$ of immersed surfaces in $X$.  But
$A[\Sigma]$ presumably does not admit such a representation, and
it is not even obvious that the space $\CM$, which should play the
role of the complex manifold $M$ in the finite-dimensional model,
admits a complex structure.

Thus, as far as we know, the full $D1$-brane worldvolume action
does not fit into the simple structure of the finite-dimensional
model. As a result, we cannot hope to use the $D1$-brane formalism
to prove vanishing results such as \corollary.  Physically, the
difficulty in using the $D1$-brane formalism to prove the
vanishing results is  that the $D1$-brane worldvolume description
becomes more complicated when the brane is ``off-shell'', i.e. not
supersymmetric.  We do not believe that these off-shell
complications are really essential, but we also do not know how to
eliminate them in the $D1$-brane framework.\foot{$D5$-branes can
be put in a gauge-invariant version of this framework.}  When we
deduce these vanishing results in Section 3, we will use instead
approaches based on linear and half-linear sigma models, which are
more closely related to the finite-dimensional model.

Yet to discuss the superpotential contribution from a $D1$-brane
which wraps an isolated holomorphic curve $C$ in $X$ requires
considerably less than the full worldvolume action.  Since we
evaluate the partition function at one-loop, we only need to
discuss fluctuations about the holomorphic curve up to quadratic
order in the action. Considering the worldvolume theory only to
this order, we can nicely fit it into the framework of the
finite-dimensional model.  In particular, the second variation of
$A[\Sigma]$ away from a minimum corresponding to a holomorphic
curve indeed appears as the norm-squared of a suitable section $s$
and the contribution to the superpotential is indeed a residue.

Geometrically, the approach of working only to quadratic order in
the supersymmetric $D1$-brane action corresponds to linearizing
the space $\CM$ over the point corresponding to the given
holomorphic curve $C$ in $X$.  The linearization possesses the
requisite complex structure.

We now give a thorough discussion of how this approximation to the
$D1$-brane action fits into the framework of the
finite-dimensional model.  As we have indicated, our
identification of the supersymmetric $D1$-brane partition function
as a residue is of more conceptual than practical interest here,
not only because of the off-shell complications but also because
of the lack of compactness in the $D1$-brane approach. However, in
Section 4 we will apply similar ideas to study the superpotential
contributions from continuous families of membrane instantons in
M-theory compactifications on manifolds of $G_2$ holonomy.

To proceed, we begin with the general observation \bsv\ that whenever
a brane wraps a supersymmetric cycle, then the worldvolume theory on
the brane is automatically twisted, implying the existence of at least
one scalar supercharge.  The existence of a scalar supercharge on the
$D1$-brane worldvolume is crucial if we are to interpret the
worldvolume theory in analogy to the finite-dimensional model, with
its scalar supersymmetry generator $\delta$.

We focus our attention on the sector of the $D1$-brane worldvolume
theory describing fluctuations of the brane in $X$, as opposed to
the trivial sector describing fluctuations in $\BR^4$.  When the
$D1$-brane wraps a holomorphic curve $C$ in $X$, the worldvolume
bosons $x^i$ and $x^{\bar i}$ which describe fluctuations of the
brane in $X$ transform as coordinates on the holomorphic normal
bundle $N$ and anti-holomorphic normal bundle $\overline{N}$ of
$C$ in $X$.  The worldvolume theory also possesses fermions
$\psi_{\dot \alpha, i}$ which transform as right-moving Weyl
fermions on $\BR^4$, as indicated by the $\dot \alpha$ index, and
as coordinates on the dual bundle $N^*$ of $N$.  Equivalently,
using the hermitian metric $g_{\bar i i}$ on $X$, we can regard
these fermions as transforming in the anti-holomorphic normal
bundle $\overline{N}$.  The twisted model has two scalar
supercharges, described in detail later, which relate the
worldvolume fields $(x^{\bar i}, \psi^{\bar i}_{\dot \alpha})$.

Now, in the finite-dimensional model, the supersymmetry
transformations as well as the form of the action are determined by the
holomorphic section $s$ of $V$.  So what are the analogues of $s$ and
$V$ for the $D1$-brane?

As has already been observed in \refs{\donald, \branes}, for a
variety of supersymmetric compactifications of string and
M-theory, the supersymmetric brane configurations can be
characterized as the critical points of a ``superpotential''
$\Psi$, suitably interpreted as a function on the space of
arbitrary brane configurations.  (This idea has also been
discussed lately in a mathematical context in \leung.)  For the
$D1$-brane, if $\delta$ is the exterior derivative on the space
$\CM$ of brane configurations, then $\delta \Psi$ is a one-form
that vanishes at the point corresponding to a holomorphic curve
$C$, and moreover $\delta \Psi$ is holomorphic once we linearize
in a neighborhood of $C$. So a natural guess is to take $V$ to be
the holomorphic cotangent bundle $T^*\CM$ and $s = \delta \Psi$.

To check that this identification is correct, we must describe $\Psi$
explicitly.  For argument's sake, we start by defining $\Psi$ on
surfaces $\Sigma$ which are homologically trivial in $X$ --- although
we note that any holomorphic surface, being calibrated by the K\"ahler
form on $X$, actually resides in a nontrivial homology class.  In any case,
$\Psi(\Sigma)$ is defined for a homologically trivial surface $\Sigma$ by
\eqn\psione{ \Psi(\Sigma) = {1 \over 6} \int_B \, \Omega \,,}
where $B$ is a bounding three-cycle for $\Sigma$ and $\Omega$ is the
holomorphic three-form on $X$.  The factor of $1 \over 6$ is simply
to cancel some constants that would otherwise appear in later
formulae.  If $H_3(X,\BZ) \neq 0$, as is always the case when $X$
has complex structure moduli, then $\Psi(\Sigma)$ generally depends
on the class of $B$ and is defined only up to an additive constant.

Now, if $\Sigma$ is a surface representing a nontrivial homology class in
$X$, then a bounding three-cycle $B$ does not exist.  To define
$\Psi(\Sigma)$ in this case, for each class in $H_2(X,\BZ)$ we choose
a particular representative $\Sigma_0$.  Then, if $\Sigma$ lies in the
same class as $\Sigma_0$, a bounding three-cycle $B$ exists for
$\Sigma - \Sigma_0$.  That is, the boundary of $B$ has two components,
one of which is $\Sigma$ and the other is $\Sigma_0$, considered with
opposite orientation.  So now we set
\eqn\psitwo{ \Psi(\Sigma) - \Psi(\Sigma_0) = {1 \over 6} \int_B \, \Omega \,.}
In this case, the additive constant in $\Psi$ also depends on the
representative $\Sigma_0$ as well as the class of $B$.

The fact that $\Psi$ is only defined up to an additive constant
does not concern us, as this constant does not affect the location
of the critical points, for which $\delta \Psi = 0$.  Explicitly,
in terms of holomorphic coordinates $x^i$ on $X$, \eqn\dpsi{
\delta \Psi(\Sigma) = \ha \int_\Sigma \, \Omega_{ijk} \, \delta x^i \,
d x^j \^ d x^k \,.} So $\delta \Psi = 0$ precisely for those
surfaces $\Sigma$ on which the $(2,0)$-form $\Omega_{ijk} \, d x^j
\^ d x^k $ is equal to zero. If $\Sigma$ is holomorphic in $X$,
then any $(2,0)$-form vanishes when restricted to $\Sigma$, so
$\delta\Psi$ vanishes when $\Sigma$ is a holomorphic curve $C$.
Because $\Omega_{ijk}$ is everywhere nonzero, holomorphy of
$\Sigma$ is necessary as well as sufficient for vanishing of
$\delta \Psi$.

While $\delta\Psi$ vanishes at the point corresponding to $C$, we
also need the linear behavior near this point.  For this, we pick
local complex coordinates on $X$ consisting of a parameter $z$
that is a local complex coordinate on $C$ as well as two local
coordinates $y^i$ of the normal bundle $N$.  We write
$\epsilon_{ij}$ for $\Omega_{zij}$.  In \dpsi, we take $\delta
x^i$ to be a displacement of one of the $y^i$, since otherwise we
are not moving $\Sigma$ away from $C$ at all.  So we will write
$\delta y^i$ for $\delta x^i$.  Evaluated on $\Sigma$, we have
$dx^j\^dx^k=dz\,\^d\bar z\,(\partial_zx^j\,\partial_{\bar
z}x^k-\partial_{\bar z}x^j\partial_zx^k)$.  Because of the
antisymmetry in $j$ and $k$ (or because $\partial_{\bar z}z=0$),
we cannot set both $x^j$ and $x^k$ equal to $z$.  To linearize
$\delta \Psi$ around $C$, we set one of them, say $x^j$, to $z$,
and the other to $y^k$.  So we get
\eqn\bestdpsi{\delta\Psi=\int_C\epsilon_{ij}\,\delta y^i\bar\partial
y^j+\dots,} where the ellipses indicate that higher order terms
have been dropped. From this, we can also deduce that to quadratic
order,
\eqn\hugo{\Psi=c+{1\over2}\int_C\epsilon_{ij}\,y^i\bar\partial y^j,}
where $c$ is an integration constant.

In particular, we see from \hugo\ that when evaluated on $C$,
\eqn\ds{{{\delta^2 \Psi} \over {\delta y^j(z)
\delta y^i(z')}}\Big|_C = \epsilon_{ij} \, \partial_{\bar z}
\delta(z,z')\,,}
where, more intrinsically, $\partial_{\bar z}$ represents the
$\bar\partial$ operator acting on sections of $N$.

Since $\Psi$ functions like a superpotential, the unbroken
worldvolume supersymmetries in the linearized theory can be very
simply expressed in terms of $\Psi$. Under the twisted
supercharges $\overline{Q}_{\dot \alpha}$, the transformations of
the fields $y^i$, $y^{\bar i}$, and $\psi_{\dot \alpha, i}$ take
the usual form \eqn\DQ{\eqalign{ &\delta_{\dot \alpha} y^i = 0
\,,\quad \delta_{\dot \alpha} y^{\bar i} = \psi^{\bar i}_{\dot
\alpha} \,,\cr &\delta_{\dot \alpha} \psi_{\dot \beta, i} =
\epsilon_{\dot \alpha \dot \beta} \, {{\delta \Psi} \over {\delta
y^i}} \,.\cr}} Since $\Psi$ is holomorphic, in the sense that
$\delta \Psi / \delta y^{\bar i} = 0$, these supersymmetry
transformations satisfy $\{ \delta_{\dot \alpha}, \delta_{\dot
\beta} \} = 0$ as required. Obviously these worldvolume
supersymmetries are unbroken when $\delta \Psi / \delta y^i = 0$,
which we have already observed is the proper condition for the
$D1$-brane to be supersymmetric.  Further, taking $s = \delta
\Psi$, we see that \DQ\ represents an $N = 2$ generalization of
the supersymmetry transformations \littleQ\ in the
finite-dimensional model.

The worldvolume action which describes to leading order the
fluctuations of a $D1$-brane which wraps a holomorphic curve $C$
in $X$ takes a very simple form when written in terms of $\Psi$.
Just as for the finite-dimensional action \sone,
\eqn\dones{\eqalign{ S &= {1 \over 4} \int_C \! \epsilon^{\dot
\alpha \dot \beta} \, \delta_{\dot \beta} \left( \omega \, g^{\bar
i i} \, {{\delta \overline\Psi} \over {\delta y^{\bar i}}} \,
\psi_{\dot \alpha, i} \right)\,\cr &= \int_C \! \omega \left( \ha
g^{\bar i i} {{\delta \overline\Psi} \over {\delta y^{\bar i}}} \,
{{\delta \Psi} \over {\delta y^i}} + {1 \over 4} \epsilon^{\dot
\alpha \dot \beta} g^{\bar i i} {{D^2 \overline{\Psi}} \over {D
y^{\bar j} D y^{\bar i}}} \psi^{\bar j}_{\dot \beta} \psi_{\dot
\alpha, i} \right) \,.\cr}}
Here $D$ is the covariant derivative with respect to the metric
$g_{i \bar i}$ on $X$, and $\omega$ is the K\"ahler form on $X$ which
restricts to the volume form on $C$. We also note from \dpsi\ that
$\delta \Psi / \delta y^i$ is actually a two-form on $C$, and we
have implicitly used the induced metric to dualize
$\delta \Psi / \delta y^i$ to a scalar on $C$.

The action $S$ is to be interpreted by expanding to quadratic
order in the normal fluctuations $ y^i$ and $ y^{\bar i}$ about
the given holomorphic curve $C$, so that
\eqn\quadS{ S = \int_C \!
\omega \left( \ha g^{\bar i i} {{D^2\overline\Psi} \over {Dy^{\bar
j} Dy^{\bar i}}} {{D^2\Psi} \over {Dy^i Dy^j}}  y^{\bar j} y^j +
{1 \over 4} \epsilon^{\dot \alpha \dot \beta} g^{\bar i i} {{D^2
\overline{\Psi}} \over {D y^{\bar j} D y^{\bar i}}} \psi^{\bar
j}_{\dot \beta} \psi_{\dot \alpha, i} \right)\,.}
Using \ds, we can write $S$ more explicitly as
\eqn\quadSS{ S = \ha \int_C \! \omega \left( g_{\bar i i} \,
\partial_z y^{\bar i} \, \partial_{\bar z} y^i + \ha
\epsilon^{\dot \alpha \dot \beta} \epsilon_{\bar i \bar j} \, \psi^{\bar
j}_{\dot \beta} \partial_z \psi^{\bar i}_{\dot \alpha} \right)\,.}
In the above, we have assumed that $\Omega$ is normalized on $X$ so
that $g^{\bar j j} \bar\epsilon_{\bar j \bar i} \epsilon_{j i} = g_{\bar i
i}$.  This action is just the free action for fluctuations which we
implicitly used in the Introduction when we evaluated the partition
function.

More geometrically, we can identify the complex linear space $\CM$
describing fluctuations of the $D1$-brane about $C$ with the space
of sections of $N$.  Our formula for $S$ simply reflects the
classic fact \refs{\simons, \lawson} that, given a section $ y^i$,
the second derivative of the area functional $A[\Sigma_t]$ along
the one-parameter family of surfaces $\Sigma_t$ determined by $
y^i$, evaluated at ${\Sigma_0 = C}$, is just \eqn\dA{ {d^2 \over
{dt^2}} A[\Sigma_t]\Big|_{t=0} = \ha \, \int_C \! \omega \, \left|
\partial_{\bar z} y^i \right|^2\,,} which appears as the
bosonic term in \quadSS.  This formula indicates that holomorphic
curves are always area-minimizing in $X$, and only holomorphic
deformations of a holomorphic curve can preserve its area.

Finally, to make contact with the finite-dimensional model, we can
evaluate the partition function $Z(C)$ of a $D1$-brane wrapped on $C$
exactly as we evaluated the contribution to the finite-dimensional
integral from an isolated, non-degenerate zero of $s$ in \zp.  We find
that
\eqn\zc{\eqalign{
Z(C) &= \int_\CM {\hbox{Pfaff}\left(\overline
\partial_{E(-1)}\right)} \, d \mu \, e^{-S}\,,\cr
&= {{{\hbox{Pfaff}\left(\overline \partial_{E(-1)}\right)}(C)}
\over {\det{(\delta^2 \Psi / \delta y^j \delta y^i)}(C)}}\,.\cr}}
Here $d \mu = \CD y^i \, \CD y^{\bar i} \,
\epsilon^{\dot \alpha \dot \beta} \, \CD \psi^{\bar j}_{\dot
\beta} \, \CD \psi_{\dot \alpha, j}$ is the naive path-integral
measure, and the Pfaffian factor produced by the left-moving
bundle fermions is directly analogous to the section $g$, since
both are required for the path-integral measure to be
well-defined.  Recalling from \ds\ that $\delta^2 \Psi / \delta
y^j \delta y^i$ represents the $\overline\partial$ operator acting
on sections of $N = \CO(-1) \oplus \CO(-1)$, we see that $Z(C)$
indeed agrees with the summand in the expression \corollary.

\newsec{A Residue Theorem for the Heterotic String}

We now extend our investigation of residues in Section 2 to the
heterotic string\foot{Because the left-moving world-sheet fermions
play only an auxiliary role in our analysis, we will not need to
distinguish between the $E_8 \! \times \! E_8$ and $Spin(32)/\BZ_2$
heterotic strings.} itself.  Our goal is to prove a residue theorem,
precisely analogous to the theorems we derived in Section 2, for the
vanishing of world-sheet instanton contributions to $W$.  A very
useful tool in our analysis is the twisted version of the
heterotic world-sheet theory, as it is the twisted theory that
directly generalizes the finite-dimensional model we introduced in Section
2.  Thus, we begin this section with a short reminder of what it
means to twist \refs{\topsig\mirror\eguchi\agntI-\agntII} the heterotic
world-sheet theory, and we explain how this theory is related to the
finite-dimensional model of Section 2.

\subsec{Preliminary Remarks on Twisting}

The twisted heterotic world-sheet theory is simply a version of the
physical (untwisted) heterotic theory in which the right-moving
world-sheet fermions are assigned unconventional spins.  To describe
the twisting, we first recall that the world-sheet theory contains
complex bosons $\phi^i$ and $\phi^{\bar i} \equiv \overline{\phi^i}$
which describe sigma model maps $\Phi:\Sigma \rightarrow X$ from the
world-sheet $\Sigma$ to a K\"ahler target space $X$.  In the physical
theory, the superpartners of $\phi^i$ and $\phi^{\bar i}$ are
right-moving fermions $\psi^i$ and $\psi^{\bar i}$, which transform as
sections of the bundles $\overline{K}^\ha \otimes \Phi^*(TX)$ and
$\overline{K}^\ha \otimes \Phi^*(\overline{TX})$ respectively.  Here,
$TX$ denotes the holomorphic tangent bundle of $X$, and $\overline{K}$
denotes the anti-canonical bundle of $\Sigma$.  The anti-canonical bundle can
be explicitly described as the line-bundle of $(0,1)$ forms on
$\Sigma$, and from this description we see that $\overline{K}^\ha$ is
a right-moving spin-bundle on $\Sigma$.  Then in the twisted theory, we
simply take $\psi^i$ to transform as a section of
$\overline{K} \otimes \Phi^*(TX)$ and $\psi^{\bar i}$ to transform as
a section of $\Phi^*(\overline{TX})$.

One way to interpret the twist is that we shift the right-moving
world-sheet stress tensor $T_{\bar z \bar z}$ by \eqn\twisting{
T_{\bar z \bar z} \rightarrow \widetilde{T}_{\bar z \bar z} =
T_{\bar z \bar z} + \ha \partial_{\bar z} j_{\bar z} \,,} where
$j_{\bar z}$ is the world-sheet $U(1)$ current present in the
right-moving $N=2$ algebra.  Upon twisting, one of the two
right-moving world-sheet supersymmetry generators becomes a
nilpotent scalar $Q$, which we interpret as a $BRST$-operator on
the world-sheet. The decoupling of $Q$-trivial states from the
correlation functions of $Q$-invariant operators then greatly
simplifies the twisted theory. In particular, though the twisted
heterotic theory is not  topological, all correlation functions of
$Q$-invariant operators in the twisted theory vary holomorphically
on the world-sheet, because the twisted stress-tensor
$\widetilde{T}_{\bar z \bar z}$ is $Q$-trivial.

We now explain how the general framework of Section 2 applies to the
twisted world-sheet theory.  Instead of performing an integral over a
finite-dimensional complex manifold $M$, we now perform a
path-integral over the infinite-dimensional complex manifold $\CM$
which is the space of all sigma model maps $\Phi: \Sigma \rightarrow X$.
The world-sheet bosons $\phi^i$ themselves provide local holomorphic
coordinates on $\CM$ and play the same role as the holomorphic
coordinates $z^i$ on $M$.  In addition, the fermions $\psi^{\bar i}$,
as sections of $\Phi^*(\overline{TM})$, are coordinates on
$\overline{T \CM}$ and correspond to the anti-commuting coordinates
$\theta^{\bar i}$ in Section 2.  Finally, we interpret the
fermions\foot{We have slightly changed notation $\psi^i \rightarrow
\psi^i_{\bar z}$ to remind ourselves that $\psi^i_{\bar z}$ now
transforms as a $(0,1)$ form on $\Sigma$.} $\psi^i_{\bar z}$, which
transform as sections of $\overline{K} \otimes \Phi^*(TX)$, as
anti-commuting coordinates on a holomorphic bundle $\CV$ over $\CM$, so that
these fermions play the same role as the fermionic coordinates
$\chi^\alpha$ on $V$ in Section 2.  In particular, on world-sheets for which
$\overline{K}$ is trivial, we can identify the bundle $\CV$ on $\CM$ as
the holomorphic tangent bundle $T \CM$.

Under $Q$, the world-sheet fields transform as
\eqn\bigQ{\matrix{
&\eqalign{
&\delta \phi^i = 0\,,\cr
&\delta \psi^i_{\bar z} = \partial_{\bar z} \phi^i \,,\cr}
&\eqalign{
&\delta \phi^{\bar i} = \psi^{\bar i} \,,\cr
&\delta \psi^{\bar i} = 0 \,.\cr}\cr}}
Comparing \bigQ\ to \littleQ, we see that the action of $Q$ is
precisely analogous to the supersymmetry transformation in the
finite-dimensional model.  Further, we see that $\partial_{\bar z}
\phi^i$ is the holomorphic section of $\CV$ corresponding to the
section $s$ of $V$ in Section 2.

The sigma model action for the world-sheet fields can now be written
as
\eqn\Ssigma{ S = \int_\Sigma d^2 z \, \delta \left( g_{\bar i i} \,
\partial_z \phi^{\bar i} \, \psi^i_{\bar z} \right) \, + \, \cdots \,,}
where $g_{\bar i i}$ is the K\"ahler metric on $X$.  The
$Q$-trivial expression above is a direct generalization of the
action \sone\ which we considered in Section 2.  Just as the
finite-dimensional integral localizes on the set where $s=0$, so
the twisted path-integral localizes on sigma model maps satisfying
$\partial_{\bar z} \phi^i = 0$.  Such maps, being holomorphic, are
either constant or represent world-sheet instantons.

The ``$\cdots$'' appearing in $S$ represents the additional terms in
the sigma model action which are not $Q$-trivial
(but of course are $Q$-closed).  More precisely, these additional
terms arise either from a purely topological expression which is the
integral of the complexified K\"ahler class of $X$ over $\Sigma$ or
from the kinetic terms of the left-moving bundle fermions.

Both of these sorts of terms admit an easy interpretation in light of
the results of Section 2.  First, if we restrict the world-sheet
path-integral to the sector describing maps whose images lie in a
fixed homology class of $X$, the topological term in $S$ is constant
and can be ignored.  Second, if we also only consider world-sheet
correlation functions which do {\it not} involving the left-moving
bundle fermions, then at least for isolated world-sheet instantons, the
only role of the bundle fermions is to produce the Pfaffian factor that
appears in the Introduction.  As we have already observed in the
contex of the $D1$-brane, like the section $g$ in the finite-dimensional
model, this Pfaffian factor can be interpreted as defining a suitable
measure for the path-integral over the modes of $\phi^i$, $\phi^{\bar i}$,
$\psi^i_{\bar z}$, and $\psi^{\bar i}$.

Finally we remark that, although the physical and twisted theories
are generally very different, some quantities in the physical theory can be
computed using the twisted theory.  In particular, as long as $K$ is
the trivial bundle on $\Sigma$, correlation functions computed on
$\Sigma$ in the twisted theory agree with those computed on $\Sigma$
in the physical theory.  For instance, if $\Sigma$ is a cylinder
with Ramond sector ground-states at each end, then correlation
functions on $\Sigma$ compute the Yukawa couplings arising from the
superpotential $W$ in the low-energy effective theory.  In this
fashion we can use the twisted theory to probe for a background $W$.

\subsec{The Half-Linear Heterotic String}

Our proof of the residue theorem in Section 2 only relies upon the
fact that the integral $Z$ is invariant under a nilpotent
supersymmetry and the fact that the space $M$ over which we integrate
is compact.  We wish to generalize this residue theorem to apply to
world-sheet correlators in the twisted heterotic theory, so we
must consider models for which both of these crucial facts hold.
Since the action of the BRST-operator $Q$ on the world-sheet naturally
generalizes the supersymmetry transformation of Section 2, the first
fact holds for an arbitrary $(0,2)$ compactification.  However,
as regards the second fact, the space $\CM$ of sigma model maps
is certainly not compact, and so to generalize the finite-dimensional
residue theorem from Section 2 to a vanishing result for $W$ of the
form \corollary, we must look for heterotic models with some special
sort of compactness.

The vanishing result of \silver\ naturally suggests that we start by
considering the linear sigma models.  Indeed, the compactness of the
moduli spaces of $X$ and $E$ is an essential ingredient in the
analysis of \silver.

Moreover, the linear sigma models possess another sort of compactness
not present in an arbitrary heterotic compactification.  As discussed
extensively in \morples\ and \losev, in compactifications for which
continuous families of world-sheet instantons exist, the instanton
moduli spaces of the linear sigma model provide natural
compactifications of the instanton moduli spaces of the corresponding
nonlinear sigma model.  The compactness of these instanton moduli
spaces turns out to be the essential ingredient in our proof of a
residue theorem for the heterotic string.

However, we do not really have to consider the linear sigma models
themselves to exploit the fact that the instanton moduli spaces of the
corresponding nonlinear sigma models have natural compactifications.
We find it technically simpler, in fact, to discuss a class of half-linear
heterotic models.  These models are like the linear models in that $X$ is a
complete-intersection Calabi-Yau in a compact toric variety $Y$.
Unlike the linear models, the gauge bundle $E$ on $X$ is any bundle
which satisfies the usual consistency conditions on $X$ and also
pulls back from a bundle on $Y$.  Thus $E$ must generally be described
in a nonlinear fashion.

So in the remainder of this section, we first introduce the half-linear
models and demonstrate that the finite-dimensional residue theorems
of Section 2 naturally generalize to formulae of the form \corollary.
We then return to the linear sigma models themselves and give a direct
proof of the vanishing of instanton contributions to $W$.  For
concreteness, we shall throughout this section consider only the case
that $X$ is the quintic hypersurface in $Y = \BC \BP^4$.

\bigskip\noindent{\it Half-linear fields}\smallskip\nobreak

We start by specifying the field content of the half-linear model.
The world-sheet bosons and the right-moving world-sheet fermions are the
usual fields which describe twisted nonlinear sigma model maps
$\Phi: \Sigma \rightarrow Y$.  For the case $Y = \BC \BP^4$, the model
has four complex bosons $\phi^i$ and $\phi^{\bar i}$ which
represent {\it local} holomorphic and anti-holomorphic coordinates on
$Y$ (as opposed to the global homogeneous coordinates on $Y$ that
would appear in the corresponding linear sigma model).  Since the
half-linear model is twisted, the right-moving supersymmetry
associates to the bosons $\phi^i$ and $\phi^{\bar i}$ corresponding
fermions $\psi^i_{\bar z}$ and $\psi^{\bar i}$, transforming on
$\Sigma$ as sections of $\overline{K} \otimes \Phi^*(TY)$ and
$\Phi^*(\overline{TY})$ respectively.

As for the left-moving sector of the world-sheet, the bundle $E$
on $Y$ is represented in the usual nonlinear fashion by a set of
thirty-two left-moving fermions $\lambda^a$ coupled to the pull
back of $E$ to $\Sigma$.  We assume that $E$ satisfies the
standard topological conditions for anomaly-cancellation and
stability on $X$.  Thus,  $E$ satisfies  $p_1(E)/2 = c_2(TX)$
(and, if the structure group of $E$ reduces to a subgroup with
$U(1)$ factors, there are restrictions on the corresponding first
Chern classes).

However, the field content of the half-linear model, as it stands,
cannot be correct.  As in the linear sigma model, to localize the
half-linear model from $Y$ onto $X$, we must introduce a potential
$J(\phi^i)$ on the world-sheet.  Geometrically, $J$ transforms as a
holomorphic section of the line-bundle $\CO(5)$ on $Y$.  Supersymmetry
requires that $J$ couple to the right-moving fermions as well as the bosons,
but we currently have no way to couple $J$ to these fermions.

A more fundamental problem is that, although we choose the bundle
$E$ so as to cancel sigma model anomalies on $X$, the half-linear
model on $Y$ is currently anomalous as $p_1(E)/2\not= c_2(TY)$.

We can elegantly fix both of these problems by adding a pair of
left-moving fermions to the model.  These fermions,
which we denote by $\chi_z$ and $\overline{\chi}$, transform on the
world-sheet as sections of $K \otimes \Phi^*(\CO(-5))$ and
$\Phi^*(\CO(5))$.  Thus we can directly include the required Yukawa terms
for $J$ in the model.

As for the anomalies, since $\chi_z$ and $\overline{\chi}$ are
also ``twisted'' in the sense of having non-standard world-sheet
spins, they cancel the excess left-moving central charge from the
new boson.  Also, upon adding $\chi_z$ and $\overline{\chi}$ to
the left-moving sector of the model, we cancel the sigma model
anomalies, since near $X$, the adjunction formula implies that
$TY$ splits as a smooth bundle into the sum $TX \oplus \CO(5)$.
Explicitly, relative to the model on $X$, the half-linear model on
$Y$ has an additional pair of twisted, right-moving fermions which
arise from the directions in the normal bundle $\CO(5)$ to $X$ in
$Y$. These fermions transform as sections of $\overline{K} \otimes
\Phi^*(\CO(5))$ and $\Phi^*(\CO(-5))$.  Since $\chi_z$ and
$\overline{\chi}$ transform as the complex conjugates of these two
fermions, they cancel the corresponding anomalies.
\eject
\noindent{\it Half-linear supersymmetry}\smallskip\nobreak

In the half-linear model, the action of the scalar supercharge $Q$
slightly generalizes \bigQ, due to the transformations of the
left-moving fermions $\chi_z$ and $\overline{\chi}$ --- the other
left-moving fermions are invariant.  So $Q$ acts as
\eqn\halfQ{\matrix{
&\eqalign{
&\delta \phi^i = 0 \,,\cr
&\delta \psi^i_{\bar z} = \partial_{\bar z} \phi^i \,,\cr
&\delta \overline \chi = J(\phi^i) \,,\cr
&\delta \chi_z = 0 \,.\cr}
&\eqalign{
&\delta \phi^{\bar i} = \psi^{\bar i}\,,\cr
&\delta \psi^{\bar i} = 0\,,\cr\cr\cr}\cr}}
As we have mentioned, $J$ is locally a quintic polynomial in the holomorphic
coordinates $\phi^i$ and globally a holomorphic section of $\CO(5)$ on
$Y$.  Of course, $J$ represents the data needed to determine $X$ as a
hypersurface in $Y$.

We see from the action of $Q$ that the fermions $\psi^i_{\bar z}$,
$\overline{\chi}$, and $\chi_z$ in the half-linear model can all
be identified as the analogues of the fermions $\chi^\alpha$ in
the finite-dimensional model.  In this basis, $s=(\partial_{\bar
z} \phi^i,J(\phi^i),0)$.  So, if we construct an action for the
half-linear model analogous to \sone\ in the finite-dimensional
model, the half-linear model will localize on sigma model maps
$\Phi$ satisfying 
\eqn\wsinst{ \partial_{\bar z} \phi^i = J(\phi^i) = 0\,.}
The first condition requires that $\Phi$ be holomorphic,
and the second condition requires that the image of $\Phi$ lie in the
subset $J=0$ of $Y$, which can be identified with $X$.  So the half-linear
model localizes on world-sheet instantons in $X$.

\bigskip\noindent{\it The half-linear action}\smallskip\nobreak

To complete our description of the half-linear model, we must finally
specify its world-sheet action $S$.

First, in complete analogy to the action of the finite-dimensional
integral, $S$ includes the terms
\eqn\szero{\eqalign{
S_0 &= t \int_\Sigma \! d^2 z \; \delta \left( g_{\bar i i} \,
\partial_z \phi^{\bar i} \psi^i_{\bar z}+
 \overline{J} \, \overline{\chi} \right) \,\cr &= t \int_\Sigma
\! d^2 z \left( g_{\bar i i} \, \partial_z \phi^{\bar i} \,
\partial_{\bar z} \phi^i + g_{\bar i i} \, D_z \psi^{\bar i} \,
\psi^i_{\bar z}+\overline{J} J + \psi^{\bar i} D_{\bar i}
\overline{J} \, \overline{\chi} \right)\,. \cr}} Here $t$ is a
coupling parameter as in Section 2, and $g_{\bar i i}$ is a
K\"ahler metric on $Y$.  Because $S_0$ is $Q$-exact, quantities
which we compute in the half-linear model are unchanged under
deformations of $t$, $g_{\bar i i}$, and $\overline{J}$.  We also
observe parenthetically that, since the expression $\overline{J}
\, \overline{\chi}$ transforms as a smooth section of the trivial
bundle on $Y$, we do not actually need to specify a hermitian
bundle metric on $\CO(5)$ to make sense of this expression.

The action $S_E$ for the left-moving fermions $\lambda^a$ which describe $E$
is the standard action, which we record for completeness below,
\eqn\sleft{
S_E = \int_\Sigma \! d^2 z \, \left( \lambda_a D_{\bar z} \lambda^a
+ F_{\bar i i b}^{\; \; a} \, \lambda_a \lambda^b \psi^{\bar i}
\psi^i_{\bar z} \right)\,,}
where
\eqn\bundled{
\lambda_a D_{\bar z} \lambda^a = \lambda_a \partial_{\bar z} \lambda^a
+ \lambda_a \, \partial_{\bar z} \phi^i A_{i b}^{\, a} \,
\lambda^b \,.}
In the above, $A_{i b}^{\, a}$ is a holomorphic connection on $E$,
having components only of type $(1,0)$ on $Y$, and
$F_{\bar i i b}^{\; \; a}$ is the curvature of this connection.  Since
$S_E$ is the usual action for the left-moving bundle fermions, and
since $Q$ acts in the usual way \bigQ\ on the fields
appearing in $S_E$, this action is clearly $Q$-invariant.

A more nontrivial fact is that we can also write a $Q$-invariant action for
the fermions $\chi_z$ and $\overline{\chi}$.  Abstractly, the presence
of $\chi_z$ and $\overline{\chi}$ in the half-linear model implies
that we are dealing with the over-determined case $\dim M < \rk V$
discussed in Section 2.3.  So we must add some $Q$-invariant observable
$\CO$ involving $\chi_z$ and $\overline{\chi}$ to the action if we wish to
compute something nontrivial in the half-linear model.

Physically, this $Q$-invariant observable $\CO$ must introduce a kinetic term
$D_{\bar z} \overline{\chi} \, \chi_z$ for $\chi_z$ and
$\overline{\chi}$.  To find a $Q$-invariant extension of this kinetic
term, we follow the philosophy of Section 2.3 and attempt to solve
$\delta \CO = 0$ perturbatively.  We begin by noting that the
expression $\CO^{(0)}$,
\eqn\ozero{\eqalign{
\CO^{(0)} &= D_{\bar z} \overline{\chi} \, \chi_z
- \psi^i_{\bar z} \, D_i J \, \chi_z \,\cr
&= \left( \partial_{\bar z} + \partial_{\bar z} \phi^i A_i \right)
\overline{\chi} \, \chi_z - \psi^i_{\bar z} \left( \partial_i
+ A_i \right) \! J \, \chi_z \,,\cr}}
is trivially invariant under the variations of $\overline{\chi}$ and
$\psi^i_{\bar z}$.  That is, in analogy to the finite-dimensional
model, $\iota_s \CO^{(0)} = 0$.

In the expression for $\CO^{(0)}$ above, we have introduced the
canonical holomorphic connection $A_i$ on the line-bundle $\CO(5)$ on
$Y$.  Because $A_i$ depends on $\phi^{\bar i}$ as well as
$\phi^i$, we have that
\eqn\varyco{ \delta \CO^{(0)} = \overline \partial \CO^{(0)} \neq 0 \,.}
Rather,
\eqn\delozero{
\overline \partial \CO^{(0)} =
F_{\bar i i} \, \psi^{\bar i} \, \partial_{\bar z} \phi^i \,
\overline{\chi} \, \chi_z - F_{\bar i i} \, \psi^{\bar i}
\psi^i_{\bar z} \; J \, \chi_z \,,}
where $F_{\bar i i}$ is the curvature of $A_i$.  However, introducing
$\CO^{(1)}$,
\eqn\oone{
\CO^{(1)} = F_{\bar i i} \, \psi^{\bar i} \psi^i_{\bar z} \,
\overline{\chi} \, \chi_z \,,}
we easily see that
\eqn\correct{ \overline \partial \CO^{(0)} + \iota_s \CO^{(1)} = 0\,.}
Because $A_i$ is a holomorphic connection on $\CO(5)$, the curvature
satisfies $\overline \partial F_{\bar i i} = 0$, so that
$\overline \partial \CO^{(1)} = 0$.  Consequently,
$\CO = \CO^{(0)} + \CO^{(1)}$ is $Q$-invariant (but not $Q$-trivial).

Thus, we can add kinetic terms for $\chi_z$ and $\overline \chi$ to
the action $S$ by including
\eqn\schi{\eqalign{
S_\chi &= \int_\Sigma \! d^2 z \, \CO = \int_\Sigma \! d^2 z \, \left(
\CO^{(0)} + \CO^{(1)} \right) \,\cr
&= \int_\Sigma \! d^2 z \, \left( D_{\bar z} \overline{\chi} \, \chi_z
- \psi^i_{\bar z} \, D_i J \, \chi_z + F_{\bar i i} \, \psi^{\bar i}
\psi^i_{\bar z} \, \overline{\chi} \, \chi_z \right) \,.}}

Finally, we include in $S$ the purely topological term which describes
the action of the world-sheet instanton itself,
\eqn\stop{ S_{top} = \int_\Sigma \Phi^*( \omega_\BC )\,,}
where $\omega_\BC$ is the complexified K\"ahler class of $Y$.  This
term simply reproduces the exponential factor in \holwcinst, but we
include it for completeness.

Thus, the action for the half-linear model is
\eqn\bigS{ S = S_0 + S_E + S_\chi + S_{top}\,.}

\subsec{A Half-Linear Residue Theorem}

We now show in the half-linear model that world-sheet instanton
contributions to the superpotential $W$ vanish by a residue theorem
precisely analogous to the finite-dimensional residue theorem of
Section 2.

Before we discuss a residue theorem for the half-linear model, though,
we must first demonstrate the general fact that the half-linear model on
$Y$ is equivalent to the usual twisted non-linear sigma model on
$X$.  Only then does the residue theorem for the half-linear model
imply the vanishing of the instanton contributions to $W$ in the
non-linear sigma model.

Relative to the non-linear model on $X$, the half-linear model on $Y$
possesses additional world-sheet degrees of freedom described by the
left-moving fermions $\chi_z$, $\overline{\chi}$, and the complex
boson and associated right-moving fermions describing fluctuations of
the world-sheet normal to $X$ in $Y$.  We will denote these normal
fields simply by $\phi$, $\overline{\phi}$, $\psi_{\bar z}$,
and $\overline{\psi}$, suppressing indices associated to the tangent
bundle $TY$.

The additional world-sheet fields present in the half-linear model on
$Y$ relative to the non-linear model on $X$ are all massive due to the
terms in the action involving $J$ and $\overline{J}$.  For instance, the
normal bosons $\phi$ and $\overline{\phi}$ gain a mass from the
$\overline{J} J$ term that appears in the $Q$-trivial action $S_0$,
\eqn\zeromodes{t \int_\Sigma \! d^2 z \,
\delta \left( \overline{J} \overline{\chi} \right) =
t \int_\Sigma \! d^2 z \, \left( \overline{J} J +
\psi^{\bar i} D_{\bar i} \overline{J} \, \overline{\chi} \right) \,.}
Similarly, the fermions $\psi_{\bar z}$, $\overline{\psi}$, $\chi_z$,
and $\overline{\chi}$ all gain masses from the $\overline{DJ}$ term in
\zeromodes\ and the conjugate term appearing in $S_\chi$ in \schi.

The mass terms for $\phi$, $\overline{\phi}$, $\psi_{\bar z}$,
$\overline{\psi}$, $\chi_z$, and $\overline{\chi}$ thus appear in the
half-linear action as
\eqn\masses{ \int_\Sigma \! d^2 z \, \left( \overline{\phi} \,
\overline{DJ} DJ \, \phi + \overline{\psi} \, \overline{DJ} \, \overline{\chi}
- \psi_{\bar z} \, DJ \, \chi_z \right)\,,}
where $DJ$ is the (holomorphic) normal derivative of $J$ along $X$,
and we have absorbed the coupling $t$ in \zeromodes\ into
$\overline{J}$.  Because we assume that $X$ is a non-singular quintic
hypersurface, $DJ$ is everywhere non-vanishing on $X$ and consequently
transforms in the trivial line-bundle on $X$.  Also, because $J$ is
holomorphic and vanishes on $X$, the $\overline\partial$ operator of
$X$ acts on $DJ$ as $\overline\partial DJ = [\overline\partial,D] J =
0$, so that $DJ$ is holomorphic on $X$.  As such, once we choose
a non-vanishing holomorphic section of the trivial bundle on $X$, a
choice which we must make in defining the fermionic measure of
the path-integral, we can regard $DJ$ as merely a constant mass
parameter for the normal modes.

As we have already remarked, since $\overline{J}$ only appears in
the half-linear model through the $Q$-trivial terms in \zeromodes,
the half-linear model is invariant under deformations of
$\overline{J}$.  Scaling $\overline{J}$ by a large constant, the
massive world-volume fields in \masses\ all acquire arbitrarily
large masses.\foot{Looking back at \masses, the $\overline{DJ} DJ$
term is a mass for scalars of order $\sqrt{|\bar J|}$.  The
$\overline{\psi} \, \overline{D J} \, \bar\chi$ term is part of a
fermion mass matrix $\left(\matrix{0 & DJ \cr \bar{DJ} & 0
\cr}\right)$ where the upper right entry comes from the $\psi DJ
\chi$ term; it again leads to masses of order $\sqrt{|\bar J|}$.}
As such, we can integrate out these massive world-sheet fields at
one-loop with arbitrary precision.  From \masses, we see that the
one-loop contributions from the massive modes of $\phi$,
$\overline{\phi}$, $\psi_{\bar z}$, $\overline{\psi}$, $\chi_z$,
and $\overline{\chi}$ all cancel but for a finite, anomalous
factor associated to the index of the $\overline\partial$ operator
acting on the pull back of the normal bundle $N$ to the
world-sheet.  This one-loop contribution can be absorbed into a
renormalization of the string coupling constant and the K\"ahler
class of $X$ and is not relevant for the vanishing argument.
Finally, upon integrating out the massive fields, we set them to
zero in the half-linear action and in all observables, so that the
half-linear model on $Y$ clearly localizes to the non-linear model
on $X$.

For completeness, we give in this paragraph a brief description of
the renormalization.  Massive modes with nonzero momentum cancel
in the path integral, so the renormalization comes from the
constant modes.   The constant modes contribute a factor
\eqn\oneloop{ {1 \over {(\overline {D  J} DJ)^{n_1}}} \,
\overline{DJ}^{n_2} \, DJ^{n_3}\,,} where $n_1$ formally denotes
the number of modes of $\phi$, $n_2$ the number of modes of
$\overline{\chi}$, and $n_3$ the number of modes of $\chi_z$.
Since $\phi$ and $\overline{\chi}$ both transform in the pull back
$\Phi^*(N)$ of the normal bundle $N = \CO(5)$ to the world-sheet,
$n_1$ equals $n_2$.  However, $\chi_z$ transforms in the bundle $K
\otimes \Phi^*(N^*)$, and thus the difference $n_1 - n_3$ is equal
to the index of the $\overline\partial$ operator on the
world-sheet acting on the bundle $\Phi^*(N)$.  If the world-sheet
is a Riemann surface $\Sigma$ of genus $g$ and $\Phi$ is a map of
degree $d$ into $X$, then the index theorem (or simply the
Riemann-Roch theorem) implies that $n_1 - n_3 = 5d + 1 - g$. Thus,
in this situation the one-loop contribution of the massive modes
is a factor $(1 / DJ)^{(5d + 1 - g)}$.  In the non-linear model on
$X$, this one-loop contribution can be written as \eqn\newtop{
\left( {1 \over {DJ}} \right)^{(5d+1-g)} = \exp{\left[ -{1 \over
{2 \pi}} \int_\Sigma \! \log{(DJ)} \, \left( 5 \Phi^*(\omega) 
+ \ha R \right)\right]}\,,} where $\Phi^*(\omega)$ is the pull back of
the K\"ahler class from $X$, which we assume is normalized to
satisfy $\int_\Sigma \Phi^*(\omega) = 2 \pi d$, and $R$ is the
world-sheet curvature, which satisfies $\int_\Sigma R = 4 \pi (1
- g)$. The expression in \newtop\ manifestly represents the 
renormalization of the K\"ahler class of $X$ and the string coupling 
constant upon integrating out the massive modes.

Having shown that the half-linear model on $Y$ is equivalent to the
non-linear sigma model on $X$, we now establish a residue theorem for
the half-linear model which implies the vanishing of world-sheet
instanton contributions to the superpotential $W$.

The half-linear model is a closer cousin to the usual world-sheet
CFT description of the heterotic string than to the dual
$D1$-brane description which we explored in Section 2.4.  As such,
in neither the world-sheet CFT nor the half-linear model can we
compute $W$ directly.  Rather, because of the presence of three
left- and right-moving fermion zero-modes arising from
fluctuations tangent to the world-sheet, we must indirectly probe
for $W$ by computing a cubic correlator of vertex operators on the
world-sheet.  In the terminology of Section 2.3, the half-linear
model describes the over-determined case $\rk \CV > \dim \CM$, due
to the presence of the fermions $\chi_z$ and $\overline{\chi}$ in
the model, but the section $s=(\partial_{\bar
z}\phi^i,J(\phi^i),0)$ still vanishes over a locus on $\CM$ of
complex dimension three, due to the $SL(2,\BC)$ action on the
world-sheet.  So we must insert a suitable observable $\CO$, the
cubic correlator of vertex operators, to compute something
non-trivial.

The easiest way to probe for $W$ is to compute the correlator
$\langle R R R \rangle$, where $R$ is the vertex operator for the
(unique) K\"ahler modulus of $Y$.  Explicitly, \eqn\bigR{ R =
\omega_{\bar i i} \partial_z \phi^i \psi^{\bar i} \,,} where
$\omega_{\bar i i}$ is a harmonic representative of the K\"ahler
class of $Y$, implying that $R$ is $Q$-invariant.  Since the
half-linear model arises from a sigma model on $Y$ (and only
restricts to $X$ when $\overline{J}$ is large), we must consider
operators such as $R$ which are actually defined on $Y$.  Note as
claimed that each of the three right-moving fermion zero-modes
from $SL(2,\BC)$ can be soaked up with the fermion $\psi^{\bar i}$
that appears in $R$.

Of course, the K\"ahler class of $Y$ determines by restriction the K\"ahler
class of $X$ and thus the radius of the compactification.  The only
dependence of $W$ on this K\"ahler modulus is through the exponential
factor $\exp{\left( - \int_\Sigma \, \Phi^*(\omega) \right)}$ arising
from the classical action of the instanton itself.  If we let $\CR$ be
the $\CN=1$ chiral field in the low-energy effective theory associated
to the K\"ahler modulus, then the correlator $\langle R R R \rangle$
computes the third derivative $\partial^3_\CR W$ of $W$ with respect
to $\CR$.  Thus, given the simple exponential dependence of $W$ on
$\CR$, the vanishing of $W$ is equivalent to the vanishing of the
correlator $\langle R R R \rangle$.

In the case of the finite-dimensional model in Section 2, we deduced a
residue theorem by taking $t=0$.  Although we have already
interpreted the half-linear model as being formally analogous to the
finite-dimensional model, unlike the case of the finite-dimensional
model, we cannot simply take $t=0$ in the half-linear model to deduce
that $\langle R R R \rangle$ vanishes.  Clearly with no
exponential suppression of the fluctuating modes in the half-linear
model, the half-linear path-integral ceases to be defined.

However, in localizing the half-linear model on $Y$ to the non-linear
model on $X$, we have already used the fact that the $BRST$-invariance of
the half-linear model implies that the model is formally independent
of $\overline{J}$ as well as $t$.  So rather than taking $t=0$, we
consider taking $\overline{J} = 0$ instead.

When $\overline{J} = 0$, the half-linear model no longer localizes on
instantons contained in $X$.  Instead, after integrating out at weak
coupling all fluctuating modes of the fields, the half-linear model
localizes onto the moduli space of instantons in $Y$.

If we restrict attention to a given instanton sector of degree $d$
holomorphic maps $\Phi$ from $\Sigma = \BC \BP^1$ to $Y$, then the
moduli space of these instantons has a natural compactification to
$\BC \BP^{5d+4}$. Because of this compactness, the half-linear
path-integral over each instanton sector can be defined even when
$\overline{J} = 0$. Thus, the correlator $\langle R R R \rangle$
can be computed either at large $\overline{J}$, where it is
proportional to $W$ as computed in the non-linear model on $X$, or
it can be computed at $\overline{J} = 0$, where we will easily see
that it vanishes order by order in $d$.  Morally speaking, the
vanishing of the instanton contribution to the superpotential
follows by applying the residue theorem of section 2 to the
compact manifold $\BC\BP^{5d+4}$.  Rather than invoking this theorem
(which could lead one to worry about singularities in
$\BC\BP^{5d+4}$), we will imitate its proof and just look at what
happens at $\bar J=0$.

We now review in detail how $\BC \BP^{5d+4}$ arises as a
compactification of the moduli space of degree $d$ instantons in $Y$,
following \refs{\phases, \morples, \losev}.  In fact, even though we
focus here on the case $Y = \BC \BP^4$, the existence of such a
compactification generalizes whenever $Y$ is a compact toric variety,
as already applied in \refs{\morples, \losev}.

We first introduce homogeneous coordinates
${[\Phi^0:\cdots:\Phi^4]}$ on $Y$ and homogeneous coordinates $[U:V]$ on
$\Sigma$.  In terms of the homogeneous coordinates, any degree $d$
holomorphic map $\Phi: \Sigma \rightarrow Y$ is specified by a set of
homogeneous, degree $d$ polynomials $\{p^0(U,V),\ldots,p^4(U,V)\}$,
\eqn\holmaps{\eqalign{
&\Phi^0 = p^0(U,V) = a^0_0 \, U^d + a^0_1 \, U^{d-1} V + \cdots +
a^0_d \, V^d\,,\cr
&\qquad \vdots\cr
&\Phi^4 = p^4(U,V) = a^4_0 \, U^d + a^4_1 \, U^{d-1} V + \cdots +
a^4_d \, V^d\,.\cr}}

Each polynomial $p^i$ is determined by its $d+1$ coefficients
$(a^i_0,\ldots,a^i_d)$, and the space of polynomials
$\{p^0,\ldots,p^4\}$ can be parametrized by these coefficients as
$\BC^{5(d+1)}$.  Since the coordinates ${[\Phi^0:\cdots:\Phi^4]}$ are
merely homogeneous coordinates on $Y$, defined only up to scaling, an
overall scaling of $\{p^0, \ldots, p^4\}$ does not affect the map
$\Phi$.  Subtracting from $\BC^{5(d+1)}$ the point at the origin which
does not describe an actual map into $Y$ and then taking the quotient
by the overall scaling, we find the projective space $\BC
\BP^{5d+4}$.

The only subtlety in this example is that, as just observed,
${\Phi^0 = \cdots = \Phi^4 = 0}$ does not correspond to any point in
$Y$, so that the moduli space of instantons of degree $d$ on $Y$ is
actually the subset of the parameter space $\BC \BP^{5d+4}$ for which
the polynomials $p^0,\ldots,p^4$ have no common zeroes on $\Sigma$.
The polynomials which do have at least one common zero appear as an
algebraic locus of codimension four in $\BC \BP^{5d+4}$, since we
must tune one complex parameter in any four of $p^0,\ldots,p^4$ to
reach this locus.  Thus, the moduli space of ``true'' instantons in
$Y$ is a complicated but nonetheless dense, open subset of
$\BC \BP^{5d+4}$.  In particular, $\BC \BP^{5d+4}$ gives a natural
compactification of the true moduli space.

We now consider evaluating the correlator $\langle R R R \rangle$ in
the half-linear model with $\overline{J} = 0$.  In this case, if we
consider the contribution to the correlator from the topological
sector of degree $d$ world-sheet maps, we must integrate over the
moduli space of degree $d$ instantons in $Y$ described
globally above.

This integral over the instanton moduli space is actually a
supersymmetric integral just as in Section 2, since both the
world-sheet bosons and fermions possess zero-modes when $\overline{J}
= 0$.  As our global discussion above implies, the bosons $\phi^i$,
$\phi^{\bar i}$, and their superpartners $\psi^{\bar i}$ all have
$5d+4$ zero-modes.  Of these $5d+4$ zero-modes, three zero-modes arise
from the $SL(2,C)$ action on $\BC \BP^1$ and are immediately soaked up
by the cubic correlator.  The other $5d+1$ zero-modes represent the
non-trivial holomorphic deformations of degree $d$ rational curves in
$Y$.  The left-moving fermion $\overline{\chi}$ also has $5d+1$ zero-modes,
which arise from holomorphic sections of the bundle
$\Phi^*\left(\CO(5)\right) = \CO(5d)$.  Neither the right-moving
fermions $\psi^i_{\bar z}$ nor the left-moving fermion $\chi_z$ have
any zero-modes in the instanton background.

When $\overline{J}$ is non-vanishing, these $5d+1$ interesting modes
of $\phi^i$, $\phi^{\bar i}$, $\psi^{\bar i}$, and $\overline{\chi}$
enter the half-linear model action through the $Q$-trivial terms involving
$\overline{J}$ in \zeromodes\ and through the four-fermion
interactions in \sleft\ and \schi.  In the weak coupling limit $t
\rightarrow \infty$, the four-fermion interactions are irrelevant,
since they always involve the fermions $\psi^i_{\bar z}$ which have no
zero-modes.  So the only way to absorb the zero-modes of $\psi^{\bar
i}$ and $\overline{\chi}$ is through the quadratic mass terms that
arise from $\overline{J}$.

In fact, if we consider integrating out all of the fluctuating modes
at weak coupling, to reduce the half-linear path-integral to a
finite-dimensional supersymmetric integral over these $5d+1$ modes,
then the $Q$-trivial terms involving $\overline{J}$ in \zeromodes\
implicitly represent the same finite-dimensional action \stwo\ which
we considered in Section 2.  In this case, the modes of $\psi^{\bar
i}$ represent the fermionic coordinates $\theta^{\bar i}$, the modes
of $\overline{\chi}$ represent the bundle fermions $\chi^\alpha$, and
$J$ implicitly determines a holomorphic section $s$ of a rank $5d+1$
bundle over the moduli space of instantons in $Y$ which vanishes
precisely over those instantons contained in $X$.

Just as in Section 2, once we set $\overline{J}$ to zero, then the $5d+1$
fermion zero-modes of $\psi^{\bar i}$ and $\overline{\chi}$ cannot be
absorbed when computing the correlator $\langle R R R \rangle$.
Hence, $\langle R R R \rangle$ vanishes order by order for
each sector of degree $d$ maps.  Finally, since our vanishing result
follows exactly as the residue theorem in Section 2, we naturally interpret
it as a residue theorem for instanton contributions to $W$.

\subsec{Extension to the Linear Sigma Model}

Just as in the finite-dimensional case, the argument for the
vanishing of the instanton contributions to $W$ in the half-linear
model relies only upon the right-moving world-sheet
supersymmetries and suitable compactness.  These ingredients are
also present in the $(0,2)$ linear sigma models themselves, so we
should also be able to give a similar, direct argument for the
vanishing of instanton contributions to $W$ in these models.  The
reason for doing so is that the linear sigma model version of the
argument applies to a somewhat different class of models --
bundles constructed in a simple way from polynomials, but which do
not necessarily extend over $Y={\bf CP}^4$.

We now present just such an argument.  Although the gist of the
vanishing argument for the linear sigma model is exactly the same
as for the half-linear model, we must present the details of the
argument in a slightly different way, since the specifics of the
linear model and the half-linear model are very different.
Nonetheless, the fact that the general argument does extend from
the half-linear to the linear model, despite the obvious
differences between these world-sheet theories, indicates that
this argument is robust.

As in the previous section, we once more focus on the case that
$X$ is a quintic hypersurface in $Y = \BC \BP^4$.  However, we now
assume that the bundle $E$ on $X$ is a deformation of the
holomorphic tangent bundle $TX$, which corresponds in the linear
sigma model to a deformation away from the locus of theories with
$(2,2)$ world-sheet supersymmetry.  Since neither $TX$ nor $E$
pulls back (in any obvious way) from any bundle on $Y$, the
compactifications cannot necessarily be described by the
half-linear model.

\bigskip\noindent{\it Background}\smallskip\nobreak

We must recall a few facts about the $(0,2)$ linear sigma model which
describes heterotic compactification on $X$ with gauge bundle $E$.
Useful background can be found in \silver, \phases, and \distnotes.
We will be rather brief in our description of the linear sigma model,
both because this material is well-known and also because the
vanishing argument which we present does not rely on many details of
the model.

We first recall the field content\foot{We ignore the decoupled current
algebra degrees of freedom which represent the unbroken space-time
gauge group.} for this model.  On the $(2,2)$ locus itself, the linear
sigma model which describes a quintic $X$ in $\BC \BP^4$ is a
two-dimensional $U(1)$ gauge theory with five chiral superfields
$S^i$, $i=1,\ldots,5$, of charge $+1$ and one chiral superfield $P$ of
charge $-5$.

Once this model is deformed away from the $(2,2)$ locus, the $(2,2)$
gauge multiplet decomposes into a $(0,2)$ gauge multiplet and a
neutral $(0,2)$ chiral multiplet.  Similarly, each $(2,2)$ chiral
multiplet decomposes into a $(0,2)$ chiral multiplet and a $(0,2)$
Fermi multiplet.  We denote again by $S^i$ and $P$ the corresponding
$(0,2)$ chiral superfields, with components $(s^i, \psi^i_+)$ and
$(p, \psi^0_+)$, and by $\Psi^i_-$ and $\Psi^0$ the associated Fermi
superfields, with components $\psi^i_-$ and $\psi^0_-$.

The action of the $(0,2)$ model contains many interactions, but the
only interactions relevant to our vanishing argument arise from the
$(0,2)$ superpotential.  Recall that these interactions can be written
as integrals over half of $(0,2)$ superspace, in the form
\eqn\super{ S_J = {1 \over \sqrt{2}} \,
\int_\Sigma d \theta^+ \, \Big( \Psi^i_- J_i + \Psi^0_- J_0 \Big) +
\hbox{h.c.}}
In general, $J_i$ and $J_0$ are holomorphic functions of the chiral
fields $S^i$ and $P$.  More specifically, $J_i$ and $J_0$ take the
form
\eqn\js{\eqalign{
J_i &= P \, \Big( {{\partial F} \over {\partial S^i}} + F_i \Big),
\quad i=1,\ldots,5\,,\cr
J_0 &= F \,,}}
where $F = F(S^i)$ is a quintic polynomial in the $S^i$ which
determines $X$ as a hypersurface in $Y$, and the $F_i$ are quartic
polynomials in the $S^i$ that are assumed to satisfy $S^i F_i = 0$
and which determine $E$ as a deformation of $TX$.

In terms of the component fields, the $(0,2)$ superpotential \super\
leads to a bosonic potential $U$,
\eqn\boseU{ U = \sum_{i=0}^5 \, |J_i|^2 \,,}
and Yukawa interactions of the form
\eqn\yuks{ \psi_-^i \psi_+^j \, {{\partial J_i} \over {\partial S^j}}
+ \psi_-^i \psi_+^0 \, {{\partial J_i} \over {\partial P}}
+ \hbox{h.c.}}

The superpotential \super\ also preserves a right-moving $U(1)$
$\CR$-symmetry, under which the lowest components of $S^i$ and
$\Psi_-^i$ carry charge $+{1 \over 5}$, and the lowest components of $P$
and $\Psi_-^0$ are neutral.

\bigskip\noindent{\it The vanishing theorem}\smallskip\nobreak

The first step in our vanishing argument is to twist the $(0,2)$
linear sigma model so that the supersymmetry generator usually denoted
$\overline{Q}_+$ becomes a scalar, exactly as described in \silver.
Under this twisting, the world-sheet spin of each field is shifted
by $- \ha J_R + {1 \over 10} Q$, where $J_R$ is the $\CR$-symmetry
generator and $Q$ is the gauge-symmetry generator.  Since the gauge current
corresponding to $Q$ is of the form $\{ \overline{Q}_+, \ldots \}$, the
fact that ${1 \over 10} Q$ appears in the twist is irrelevant and is
merely for convenience, so that upon twisting all fields have integral
or half-integral world-sheet spins.

Upon twisting the model, the spins of the bosons $s^i$ are unaffected,
but the boson $p$ now has spin $-\ha$ and transforms as a section of
$K^\ha \otimes \CL^{-5}$ on $\Sigma = \BC \BP^1$.  Here $K$ is the canonical
bundle on $\Sigma$ as earlier, and $\CL = \CO(d)$ is the line-bundle
on $\Sigma$ associated to a degree $d$ instanton configuration in the
gauge field.  Also, just as in the half-twisted model, the fermions
$\psi^i_+$ and $\psi_+^{\bar i}$, for $i=1,\ldots,5$, now have spins
$+1$ and $0$ and transform as sections of $\overline{K} \otimes \CL$ and
$\overline{\CL}$.  Finally, the left-moving fermions $\psi_-^i$ and
$\psi_-^{\bar i}$ are unaffected by the twisting and transform as
sections of $K^\ha \otimes \CL$ and $K^\ha \otimes \overline{\CL}$.

To proceed with the argument, we must compute the linear sigma model
correlator analogous to $\langle R R R \rangle$ in the half-linear
model.  As explained in \silver, the linear sigma model representative
of the vertex operator $R$ describing deformations of the
K\"ahler class of $Y$ (hence also $X$) is $\lambda_-$, the left-moving
gaugino.  This fact can be motivated by observing that, since the
K\"ahler class of $Y$ is represented in the linear model by a
Fayet-Iliopolous $D$-term, the linear sigma model representative for
$R$ must come from the $(0,2)$ gauge multiplet.  The supersymmetry and
$\CR$-symmetry then determine this representative to be $\lambda_-$.
So we must compute the instanton contributions to
$\langle \lambda_- \lambda_- \lambda_- \rangle$ in the linear sigma model.

As in the half-linear model, the twisted linear model is formally
invariant under deformations of $\overline{J}_i$, $i=0,\ldots,5$,
so we consider taking $\overline{J}_i = 0$.  At first glance, one
might worry that this deformation would be singular in the linear
model, since at least in the untwisted theory, the boson $p$ has an
unbounded zero-mode which only receives a mass from the potential term
$U$ in \boseU.  However, because $p$ has spin $-\ha$ in the twisted
theory, this dangerous zero-mode is not present.  This observation was also
central to the vanishing argument of \silver, so we certainly expect
it to play a role in our argument as well.  Thus, we can compute
$\langle \lambda_- \lambda_- \lambda_- \rangle$ in the theory with
$\overline{J}_i = 0$, provided we perform the twist.

In the half-linear model, once we performed the analogous
deformation by taking $\overline{J} = 0$, we easily saw that the
correlator $\langle R R R \rangle$ vanished due to the presence of
excess fermion zero-modes which could no longer be absorbed through
world-sheet interactions.  We will now argue that the correlator
$\langle \lambda_- \lambda_- \lambda_- \rangle$ vanishes when
$\overline{J}_i = 0$ in the linear sigma model, again due to excess
fermion zero-modes.

The relevant zero-modes arise from the fermions $\psi^{\bar i}_+$ and
$\psi^i_-$, for $i=1,\ldots,5$.  In the background of a degree $d$
instanton, each fermion $\psi^{\bar i}_+$ has $d+1$
zero-modes, and each fermion $\psi^i_-$ has $d$ zero-modes (and the
conjugate partners of these fermions have no zero-modes).

To show that these fermion zero-modes cannot be absorbed in computing
the correlator $\langle \lambda_- \lambda_- \lambda_- \rangle$ with
$\overline{J}_i = 0$, we first make a few general remarks about
the computation of $\langle \lambda_- \lambda_- \lambda_- \rangle$
even when the $\overline{J}_i$ are not assumed to vanish.  First,
since all kinetic terms in the linear model are
$\overline{Q}_+$-trivial, we can by a field rescaling assume that the
couplings appearing in $J_i$ and $\overline{J}_i$ are arbitrarily
small.  Hence we can compute $\langle \lambda_- \lambda_- \lambda_-
\rangle$ perturbatively in $J_i$ and $\overline{J}_i$.

As a special case of our vanishing result, we now observe that
$\langle \lambda_- \lambda_- \lambda_- \rangle$ trivially vanishes
when $J_i = \overline{J}_i = 0$.  In this case, the model with no
superpotential describes, instead of $X$, the total space of the
line-bundle $\CO(-5)$ over $\BC \BP^4$.  As such, the model possesses
a classical global symmetry which rotates the fiber of this space
leaving fixed the base.  Under this symmetry, the superfields $S^i$
and $\Psi^i_-$, for $i=1,\ldots,5$, transform with charge $+1$ while
all other fields are uncharged.  In particular, the gaugino
$\lambda_-$ is uncharged, which distinguishes this global symmetry
from the $\CR$-symmetry.

The fermion zero-modes we discussed above are relevant precisely
because they cause this classical symmetry to be anomalous.  Due to
these zero-modes, regardless of the degree $d$, the path-integral
measure transforms with net charge $+5$ under this symmetry.  This anomaly
immediately implies that $\langle \lambda_- \lambda_- \lambda_-
\rangle$ vanishes in the theory with no superpotential.  For instance,
computing $\langle \lambda_- \lambda_- \lambda_- \rangle$
perturbatively at weak coupling, all interactions respect the
classical symmetry and so there is no way to absorb the fermion
zero-modes by pulling down fermion interaction terms from the action.
This fact is why the detailed structure of the linear model is largely
irrelevant for our argument.

We now consider the general case that $J_i$ and $\overline{J}_i$ are
non-zero.  Since the superpotential breaks the classical symmetry we
used above, the fermion Yukawa terms involving $J_i$ and
$\overline{J}_i$ in \yuks\ are candidates to soak up the zero-modes of
$\psi^{\bar i}_+$ and $\psi^i_-$ above.  However, whatever interaction
terms we bring down from the action to soak up the fermion zero-modes,
the anomaly implies that these terms must carry net charge $-5$ to
cancel the charge of the measure.  We now observe from \yuks\ that the
interactions involving $J_i$ all carry charge $+5$ and those involving
$\overline{J}_i$ carry charge $-5$.

Thus, when $\overline{J}_i$ vanishes, the zero-modes of the fermions 
$\psi^{\bar i}_+$ and $\psi^i_-$ cannot be absorbed, since perturbation 
theory in the $J_i$ can only bring down interactions of positive charge.  
This observation merely reflects the fact that the twisted fermions 
$\psi^{\bar i}_+$, which give rise to the anomaly, appear in the Yukawa 
couplings involving $\overline{J}_i$, not $J_i$, in \yuks.  Thus 
$\langle \lambda_- \lambda_- \lambda_- \rangle$ vanishes in an arbitrary 
degree $d$ instanton background, and instantons in the linear sigma model do 
not contribute to the space-time superpotential.

Of course, when $\overline{J}_i$ is non-zero, the linear model sums over 
individual instantons in $X$, and the contribution of each instanton should 
generically be non-zero.  Our argument is consistent with this fact, 
since insertions of the Yukawa couplings involving both $J_i$ and 
$\overline{J}_i$ can carry the proper charge to absorb the zero-modes.

This vanishing argument is at its heart very similar to the vanishing
argument of \silver\ that we reviewed in the Introduction.  A key fact
there is that $W$ transforms as a section of a
line-bundle of strictly negative curvature on the moduli space of
the low-energy effective theory.  Now in the context of the present
argument, the complex coefficients which define the quintic polynomial 
$F$ and the quartic polynomials $F_i$, and thus appear as couplings in the 
$J_i$, can be considered as projective coordinates on the moduli space 
of complex structures of $X$ and $E$.  

Our perturbative argument above can be rephrased as a selection rule for the 
dependence of the correlator $\langle \lambda_- \lambda_- \lambda_- \rangle$ 
on these coefficients.  This selection rule follows from formally assigning 
the complex coefficients appearing in $F$ and $F_i$ charge $-5$ under the 
anomalous symmetry, so that formally the $J_i$ are uncharged.  The anomaly 
implies that the correlator $\langle \lambda_- \lambda_- \lambda_- \rangle$, 
as a function of these coefficients, transforms homogeneously with charge 
$+5$.  As a result, the selection rule implies that 
$\langle \lambda_- \lambda_- \lambda_- \rangle$ (and hence $W$) must 
transform as a section of a line-bundle of strictly negative curvature over 
the complex structure moduli space which these 
coefficients parametrize.  In this language, our vanishing theorem follows 
simply because, when $\overline{J}_i$ vanishes, a perturbative calculation of 
$\langle \lambda_- \lambda_- \lambda_- \rangle$ in terms of the 
$J_i$ can only produce a polynomial in the complex coefficients, which 
has negative charge under the anomalous symmetry and does not have the 
required pole on the complex structure moduli space.  

\newsec{Families of Membrane Instantons}

The vanishing result which we derived for world-sheet instanton
contributions to the superpotential is a manifestation of the rigidity
inherent in holomorphic objects.  As an interesting contrast to this
result, we now consider how M-theory membranes which wrap a continuous
family of supersymmetric three-cycles in a manifold $X$ of $G_2$ holonomy
contribute to the superpotential.  The approach which we take here is
very similar to our discussion of $D1$-brane contributions to the
superpotential at the end of Section 2.  We note that the
superpotential contribution from an isolated membrane in $X$ has
already been thoroughly discussed in \hm.

Just as in the case of a $D1$-brane which wraps a holomorphic curve,
the worldvolume theory on a membrane which wraps a supersymmetric
three-cycle on $X$ is naturally twisted.  Unlike the case of the
$D1$-brane though, in the case of a supersymmetric membrane, the
sector of the worldvolume theory describing fluctuations in $X$ is
topological, as opposed to holomorphic, in character.  This fact could
hardly be otherwise, since $X$ is not a complex manifold, but it
represents a key distinction between $D1$-brane and membrane instantons.

Thus, if $\CC$ represents a continuous family of supersymmetric
membrane configurations within the space $\CM$ of all membrane
configurations in $X$, then the contribution to the superpotential
from the family $\CC$ only depends upon topological data
associated to $\CC$.  Our main result here is to show that the
contribution of the family $\CC$ to the superpotential is
proportional to the Euler character $\chi(\CC)$ of $\CC$.

Our analysis ignores singularities.  We suspect that it remain
valid even if some of the membrane instantons parametrized by
$\CC$ are singular, as long as $\CC$ itself is smooth.  A simple
example in which this is the case is that $X$ is a $T^3$-fibered
$G_2$ manifold, in which case $\CC$ is a K3 surface, of Euler
character 24.

\bigskip\noindent{\it The membrane worldvolume theory}\smallskip\nobreak

Just as in the Introduction, the most elegant way to determine the
superpotential contribution from a membrane instanton (or a family of
such instantons) is to compute the partition function of the membrane
worldvolume theory.  The structure of this theory is largely
determined by supersymmetry.  More specifically, it is determined by
the requirement that only supersymmetric membrane configurations
contribute to the partition function.  So we begin by recalling a few
facts about supersymmetric three-cycles in $X$.

To describe which three-cycles in $X$ are supersymmetric, we first
recall that $X$, as a manifold of $G_2$ holonomy, possesses a
canonical, covariantly constant three-form $\phi$.  Then, as
emphasized generally in \guk, the supersymmetric three-cycles are
those which are calibrated by $\phi$ and hence are of minimal volume
within each homology class.  That is, if $\Sigma$ is a supersymmetric
three-cycle, then the calibration condition states that on $\Sigma$,
\eqn\assc{ \phi|_\Sigma = \hbox{vol}|_\Sigma\,,}
where $\hbox{vol} = {1 \over 7} \phi \^ \* \phi$ is the volume form
associated to the metric on $X$.

Just as for supersymmetric $D1$-brane configurations, the
supersymmetric membrane configurations in $X$ can be characterized as
the critical points of a superpotential $\Psi$ on $\CM$.  $\Psi$ is
defined in a manner precisely analogous to the superpotential for
$D1$-brane configurations in a Calabi-Yau threefold.  Thus, we define
$\Psi(\Sigma)$ for any three-cycle $\Sigma$ by
\eqn\psitwo{ \Psi(\Sigma) - \Psi(\Sigma_0) = {1 \over 12}
\int_B \, \* \phi \,.}
Here $\* \phi$ is the four-form on $X$ dual to $\phi$, $\Sigma_0$ is a
fixed representative in the homology class of $\Sigma$, and $B$ is a
four-cycle bounding $\Sigma - \Sigma_0$.  Again, $\Psi(\Sigma)$ is
defined only up to an additive constant, depending on the choices of
$\Sigma_0$ and $B$.

But again, the fact that $\Psi$ is only defined up to an additive
constant does not concern us, as this constant does not affect the location of
the critical points, for which $\delta \Psi = 0$.  In terms of local
coordinates $x^i$, $i=1,\ldots,7$, on $X$,
\eqn\dpsi{ \delta \Psi(\Sigma) = {1 \over 3} \int_\Sigma \, \* \phi_{ijkl} \,
\delta x^i \, d x^j \^ d x^k \^ d x^l \,.}
Thus, $\delta \Psi(\Sigma) = 0$ when $\* \phi_{ijkl} \, d x^j \^ d x^k
\^ d x^l = 0$ on $\Sigma$.  As observed in \hl, this condition is
equivalent to the condition \assc\ that $\Sigma$ be calibrated by
$\phi$.  So the critical points of $\Psi$ correspond to supersymmetric
three-cycles in $X$.

Thus, $\delta \Psi$ is a one-form on the space $\CM$ of arbitrary membrane
configurations in $X$ which vanishes precisely over the supersymmetric
configurations.  So in this sense, $\delta \Psi$ plays much the same role
as the section $s$ we introduced in Section 2, and we expect the
action of the worldvolume theory on a supersymmetric membrane to be
expressed in terms of $\delta \Psi$, much as the action \stwo\ is expressed
in terms of $s$.

Unlike $s$, though, $\delta \Psi$ is not holomorphic, and the
space $\CM$ of membrane configurations is not even complex, even
on-shell. As a result, the supersymmetry algebra on the membrane
worldvolume takes a form slightly different from the supersymmetry
\littleQ\ considered in Section 2.

We focus on the sector of the worldvolume theory which describes
fluctuations of the membrane in $X$.  As explicitly demonstrated in
\hm, this sector is automatically twisted when the membrane wraps a
supersymmetric cycle $\Sigma$.  Normal fluctuations of the membrane in $X$ are
described on the worldvolume by four real bosons $y^i$,
$i=1,\ldots,4$, taking values in the (real) normal bundle $N$
of $\Sigma$ in $X$.  Associated to these four bosons are four fermions
$\psi^i_{\dot \alpha}$ also taking values in $N$ and transforming as
right-moving Weyl fermions in $\BR^4$, as indicated by the $\dot
\alpha$ index.

The worldvolume theory on the supersymmetric membrane then possesses two
scalar supercharges $\overline{Q}_{\dot \alpha}$.  The action of these
supercharges on the worldvolume fields can be neatly summarized by
introducing $(0|2)$ superfields $Y^i$, where
\eqn\sf{ Y^i = y^i + \theta^{\dot \alpha} \psi^i_{\dot \alpha}
+ \ha \epsilon_{\dot \alpha \dot \beta} \theta^{\dot \alpha}
\theta^{\dot \beta} F^i\,.}
In defining the superfield $Y^i$, we have introduced an auxiliary
boson $F^i$ taking values in $N$.  Even though the membrane worldvolume is
three-dimensional, the appropriate superspace is only the $(0|2)$
superspace because, just as for the $D1$-brane, we regard the bosonic
fields $y^i$ as being an infinite set of tangential coordinates to the
membrane configuration space $\CM$ at the point corresponding to a
given supersymmetric membrane configuration.

In the $(0|2)$ superspace, the action of the supercharges 
$\overline{Q}_{\dot \alpha}$ is exceedingly simple.  Namely, the supercharges
$\overline{Q}_{\dot \alpha}$ act as the fermionic derivatives
$\partial_{\dot \alpha}$, corresponding to the component transformations
\eqn\MQ{ \delta_{\dot \alpha} y^i = \psi^i_{\dot \alpha}\,,\quad
\delta_{\dot \alpha} \psi^i_{\dot \beta} = \epsilon_{\dot \alpha \dot
\beta} F^i\,,\quad \delta_{\dot \alpha} F^i = 0\,.}
We note that $\{ \overline{Q}_{\dot \alpha}, \overline{Q}_{\dot \beta}\}$ 
trivially vanishes.

The supersymmetry algebra, along with the requirement that the
membrane partition function localize on configurations for which
$\delta \Psi = 0$, determines the form of the worldvolume action on a
supersymmetric membrane.  As for the $D1$-brane, this action is really
the leading order action for fluctuations around a supersymmetric
configuration --- but given the topological nature of the membrane
worldvolume theory, the leading order action certainly suffices to
determine the partition function.

When written in terms of the $(0|2)$ superspace, the membrane
worldvolume action thus appears as
\eqn\MS{\eqalign{
S &= \int_\Sigma \! d^2 \theta \, \phi \left( \ha g_{ij}(Y) \,
\epsilon^{\dot \alpha \dot \beta} \partial_{\dot \beta} Y^i
\partial_{\dot \alpha} Y^j + \Psi(Y) \right)\,\cr
&= \int_\Sigma \! \phi \left( \ha g^{ij} {{\delta \Psi} \over {\delta
y^i}} {{\delta \Psi} \over {\delta y^j}} + 2 {{D^2 \Psi} \over {Dy^i
Dy^j}} (\psi^i \psi^j) + R_{ikjl} (\psi^i \psi^j) (\psi^k
\psi^l) \right)\,.\cr}}
In this expression, $g_{ij}$ is the metric on $X$, $R_{ikjl}$ is the 
curvature, and the canonical three-form $\phi$ appears simply to 
represent the volume-form on the supersymmetric three-cycle $\Sigma$.  We 
also note from \dpsi\ that $\delta \Psi / \delta y^i$ is actually a 
three-form on $\Sigma$, and so we have implicitly used the induced 
metric to dualize $\delta \Psi / \delta y^i$ to a scalar above.  Finally, 
we have used the shorthand $(\psi^i \psi^j)$ to indicate the $SU(2)$ singlet 
combination $\ha \epsilon^{\dot \alpha \dot \beta} \psi^i_{\dot \beta} 
\psi^j_{\dot \alpha}$, and in passing to the second line of \MS\ we 
integrated out the auxiliary bosons $F^i$.

The membrane worldvolume action \MS\ has a very familiar look.
Formally, we can interpret this action as the reduction to $0+0$
dimensions of the standard supersymmetric quantum mechanics \sumor\ on the
membrane configuration space $\CM$, with Morse function $\Psi$.  As is well
known, the partition function of supersymmetric quantum mechanics on a
finite-dimensional Riemannian manifold $M$ computes the Euler class
$\chi(M)$ of $M$.  Thus, our claim that the membrane partition
function is proportional to the Euler class $\chi(\CC)$ of the family
$\CC$ follows almost immediately now, though we still discuss
this result in detail below.

We can also compare the form of the membrane worldvolume theory to the
form of the $D1$-brane worldvolume theory (or more generally to the
holomorphic models we considered in Section 2).  Upon integrating out
the auxiliary bosons $F^i$, the supersymmetries on the membrane
worldvolume act as
\eqn\MQtwo{\eqalign{
\delta_{\dot \alpha} \, y^i &= \psi^i_{\dot \alpha} \,,\cr
\delta_{\dot \alpha} \, \psi^i_{\dot \beta} &= - \Gamma^i_{jk}
\psi^j_{\dot \alpha} \psi^k_{\dot \beta} + \ha \epsilon_{\dot \alpha
\dot \beta} g^{ij} {{\delta \Psi} \over {\delta y^j}}\,.\cr}}
In the above, $\Gamma^i_{jk}$ is the usual torsion-free affine
connection associated to the metric $g_{ij}$ on $X$; this connection
must appear so that the fermions $\psi^i_{\dot \alpha}$ transform
covariantly under reparametrizations of the $y^i$.  Comparing the
supersymmetries \MQtwo\ and action \MS\ of the membrane worldvolume
theory to the general supersymmetry \littleQ\ and action \stwo\ from
Section 2, we see that the membrane worldvolume theory is just a real,
$N=2$ version of the holomorphic models relevant for world-sheet
instantons which we considered earlier.  Clearly the one-form
$\delta \Psi$ on $\CM$ plays exactly the same role as the holomorphic
section $s$ on the complex manifold $M$, and from \MS\ we see that
at weak coupling the membrane partition function localizes on the
zeroes of $\delta \Psi$.  We also note that the $N=2$ supersymmetry
present in the membrane worldvolume theory determines a canonical
choice for the measure of the membrane partition function, as all
bosons are paired by supersymmetry with all fermions in \MQ.  So there
is no analogue here of the section $g$ which was necessary to define
a measure for the holomorphic models.

\bigskip\noindent{\it The membrane partition function}\smallskip\nobreak

Our simple description of the membrane worldvolume theory allows us to
easily evaluate the membrane partition function, even in the
degenerate case that the membranes wrap a continuous family of
supersymmetric three-cycles in $X$.

We first observe that, because $\overline{Q}_{\dot \alpha} =
\partial_{\dot \alpha}$, the worldvolume action \MS\ is evidently
$\overline{Q}_{\dot \alpha}$-trivial. As a result, the membrane
partition function $Z$ is clearly topological in character.  In
particular, $Z$ is unchanged if we multiply $\phi \rightarrow t \,
\phi$, so that taking $t$ to be large we can evaluate $Z$ at weak
coupling.  Furthermore, $Z$ is unchanged under deformations of the
metric $g_{jk}$ and even the one-form $\delta \Psi$.  This latter
observation is in clear contrast to the holomorphic models in Section
2, which were unchanged under deformations of $\overline{s}$ but
certainly depended upon $s$.

Thus, we suppose that $X$ contains a continuous family of supersymmetric
three-cycles.  Then the vanishing locus of $\delta \Psi$ on $\CM$
contains a component $\CC$ of positive dimension representing this
continuous family.  To evaluate $Z$ for membranes which wrap
three-cycles in $\CC$, we simply make a generic deformation of $\delta
\Psi$, which is small in the sense that $\delta \Psi$ still grows
sufficiently fast away from $\CC$ so that $Z$ is defined.  Under such
a deformation, we lift the degeneracy of $\delta \Psi$, which now has
a finite set of isolated zeroes on $\CC$.

At weak coupling, we can directly evaluate the contribution to $Z$ from
each non-degenerate zero of $\delta \Psi$ as a one-loop integral
over the fluctuating bosons and fermions.  Generally speaking, if $P$
is such a zero, then the contribution to $Z$ from $P$ takes the form
\eqn\bigZ{ Z_P = Z(\CN) \cdot Z(\CC)_P \,,}
where $Z(\CN)$ represents the one-loop integral over modes normal to
$\CC$, and $Z(\CC)_P$ represents the one-loop integral over the
finite number of modes tangent to $\CC$ at $P$.  Because of the topological
invariance of $Z$, the factor $Z(\CN)$ in \bigZ\ does not depend on
$P$, so that
\eqn\bigZt{ Z = Z(\CN) \cdot \sum_P Z(\CC)_P\,.}

Clearly the second factor in \bigZt\ captures the interesting
dependence of the superpotential on $\CC$.  In the Gaussian
approximation, we can express the contribution $Z(\CC)_P$ from each
point $P$ as
\eqn\critpt{ Z(\CC)_P = {{\det_\CC{\left(\partial_i \partial_j
\Psi\right)(P)}} \over {\Big|\det_\CC{\left(\partial_i \partial_j
\Psi\right)(P)}\Big|}} = \pm 1 \,,}
where the subscript $\CC$ indicates that the determinants are only
evaluated over the modes tangent to $\CC$.  Geometrically, we
recognize the expression \critpt\ as the index of the vector field
$\nabla \Psi$ (projected onto $T \CC$) at the point $P$, where it
vanishes.  Thus,
\eqn\euler{ \sum_P Z(\CC)_P = \chi(\CC)\,,}
and $Z$ is proportional to the Euler character $\chi(\CC)$ of $\CC$ as
claimed.  We could also derive this result, without explicitly
deforming $\delta \Psi$ to lift its degeneracy, by using
the four-fermion interaction in \MS\ to absorb the fermion
zero-modes tangent to $\CC$, producing the Chern-Weil representation
of the Euler character.

Finally, we remark that the factor $Z(\CN)$, studied in \hm\ for
the case of an isolated membrane instanton, is simply the formal
generalization of \critpt\ from the phase of a determinant on the
tangential directions of $\CC$ to the normal directions.  $Z(\CN)$
can thus be expressed as the sign of the Dirac operator acting on
the membrane worldvolume spinors multiplied by a factor coming
from the $C$-field. 

\bigbreak\bigskip\bigskip\centerline{{\bf Acknowledgements}}\nobreak

The work of C.B. is supported by a National Science Foundation Graduate
Fellowship and under NSF Grant PHY-9802484.  The work of E.W. is supported
in part by NSF Grant PHY-0070928.

Any opinions, findings, and conclusions or recommendations expressed in
this material are those of the authors and do not necessarily reflect
the views of the National Science Foundation.

\listrefs

\end